\documentclass[fleqn, usenatbib]{mnras}
\usepackage{newtxtext,newtxmath}
\usepackage[T1]{fontenc}
\usepackage{ae,aecompl}

\usepackage{amssymb}
\usepackage{amsmath}
\usepackage{mathtools}
\usepackage{appendix}
\usepackage{multirow}

\usepackage[usenames, dvipsnames]{color}
\usepackage[normalem]{ulem}

\definecolor{darkgreen}{rgb}{0.13, 0.55, 0.13}

\title[Clustered Supernova Feedback Tests]{Momentum Injection by Clustered Supernovae: Testing Subgrid Feedback Prescriptions}

\author[E. S. Gentry et al.]{
Eric S. Gentry,$^{1}$\thanks{E-mail: gentry.e@gmail.com}
Piero Madau$^{1}$ and Mark R. Krumholz$^{2,3}$
\\
$^{1}$Department of Astronomy and Astrophysics, University of California at Santa Cruz, 1156 High St., Santa Cruz, CA, 95064, USA\\
$^{2}$Research School of Astronomy \& Astrophysics, Australian National University, Canberra, ACT 2611, Australia\\
$^{3}$ARC Centre of Excellence for Astronomy in Three Dimensions (ASTRO-3D), Canberra, ACT Australia
}

\date{Accepted XXX. Received YYY; in original form ZZZ}

\pubyear{2019}

\begin{document}
\label{firstpage}
\pagerange{\pageref{firstpage}--\pageref{lastpage}}
\maketitle

\begin{abstract} 
Using a 1D Lagrangian code specifically designed to assess the impact of multiple, time-resolved supernovae (SNe) from a single star cluster on the surrounding medium, we test three commonly used feedback recipes: delayed cooling (e.g., used in the \texttt{GASOLINE-2} code), momentum-energy injection (a resolution-dependent transition between momentum-dominated feedback and energy-dominated feedback used, e.g., in the \texttt{FIRE-2} code), and simultaneous energy injection (e.g., used in the \texttt{EAGLE} simulations). Our work provides an intermediary test for these recipes: we analyse a setting that is more complex than the simplified scenarios for which many were designed, but one more controlled than a full galactic simulation.  In particular, we test how well these models reproduce the enhanced momentum efficiency seen for an 11 SN cluster simulated at high resolution (0.6 pc; a factor of 12 enhancement relative to the isolated SN case) when these subgrid recipes are implemented in low resolution (20 pc) runs.
We find that: 1) the delayed cooling model performs well -- resulting in 9 times the momentum efficiency of the fiducial isolated SN value -- when SNe are clustered and $10^{51}$ erg are injected per SN, while clearly over-predicting the momentum efficiency in the single SN test case;  2) the momentum-energy model always achieves good results, with a factor of 5 boost in momentum efficiency; and 3) injecting the energy from all SNe simultaneously does little to prevent over-cooling and greatly under-produces the momentum deposited by clustered SNe, resulting in a factor of 3 \emph{decrease} in momentum efficiency on the average.
\end{abstract}


%%%%%%%%%%%%%%%%%%%%%%%%%%%%%%%%%%%%%%%%%%%%%%

\section{Introduction}
\label{section:intro}

Energy and momentum injection from supernovae (SNe) are thought to be one of the key ingredients regulating galaxy formation and the thermodynamics of the interstellar and circumgalactic media. Without feedback processes that reheat and redistribute gas in galaxies, simulated galaxies are found to be too cold and too centrally compact (e.g., \citealt{1991ApJ...377..365K}). Despite its importance, however, a proper treatment of feedback in 3D hydrodynamics simulations remains elusive; at high resolutions ($\lesssim$ 7 pc for $n=1$ cm$^{-3}$ and 1 SN; \citealt{2015ApJ...802...99K}), we can simply inject $10^{51}$ erg of energy and get reasonable results, but at lower resolutions this direct injection approach yields incorrect asymptotic properties for the SN remnant (SNR), and consequently erroneous galaxy properties \citep{2018MNRAS.478..302S}.

The fundamental cause of this is a phenomenon known as over-cooling \citep{1992ApJ...391..502K}: the rate of radiative cooling in a hot astrophysical plasma is a highly non-linear function of density and temperature, and when the energy deposited by an exploding SN is spread over too large a volume or mass as a result of low resolution, this non-linearity leads to a dramatic overestimate of the cooling rate. Since the full, complex physics that describes the interaction of SN ejecta with the interstellar medium (ISM) cannot be captured at currently-realistic resolutions in large galactic and cosmological simulations, simulators have adopted a variety of simpler subgrid recipes tuned
to reproduce the ``main'' behaviour of an expanding SNR. This approach cannot hope to account for every dependence on the environment or context, so one typically starts with a simple model, and then only add new scalings as they are shown to be necessary.

One potentially strong effect is due to the clustering of core collapse SNe. Depending on the frequency of such events, it is possible that one or more SNe may occur within the remmant created by a previous explosion, giving rise to a hot, low-density cavity known as a superbubble.  Some studies have shown that feedback from superbubbles can be much more efficient than isolated SNe at ejecting mass and adding momentum to the ISM (\citealt{2013MNRAS.434.3572R,2014MNRAS.443.3463S,2014MNRAS.442.3013K,2017MNRAS.465.2471G}; see \citealt{2019arXiv190300962D} for a general discussion of regimes of SN clustering). The amount of enhancement from clustering appears to depend sensitively on the level of mixing across the interface between the hot SNR interior and the cool shell around it \citep{2019MNRAS.483.3647G, 2019arXiv190209547E} and the specific superbubble regime being studied. The turbulent mixing rate is uncertain, and likely depends on both the pre-existing clumpiness of the ISM and on the presence of magnetic fields, which suppress instabilities such as Rayleigh-Taylor that promote mixing \citep{2019MNRAS.483.3647G}. At present we lack detailed magneto-hydrodynamic simulations of SNRs including conduction with enough resolution to quantify the mixing rate, and thus the feedback boost from clustering is uncertain (see the review by \citealt{2019FrASS...6....7K}  for further discussion).

Since many traditional subgrid models for SNe do not directly account for clustering, it is important to investigate whether this could constitute a significant error in how we model feedback in galactic simulations. This is a question with at least three parts: how well does a particular subgrid recipe approximate a given superbubble, how well does it approximate each of the various regimes of a superbubble, and what is the relative occurrence frequency of each superbubble regime? For simplicity, in this paper we focus on the first question only: how well is the behaviour of a superbubble driven by 11 SNe captured by existing subgrid algorithms? We focus on this particular case because the high-resolution, state-of-the-art simulations of \citet{2017MNRAS.465.2471G} show that it has near maximal effects in terms of boosting the terminal momentum of the SNR, and thus can be used to set an upper limit on the potential error that subgrid recipes make by ignoring the effects of clustering.

In this paper, we use 1D spherically-symmetric hydrodynamic simulations to study what happens when a number of common, traditional subgrid recipes are applied to a series of clustered SNe. 
We use these models to simulate the feedback of 11 SNe at both low resolution ($\Delta r_0 = 20$ pc, typical of isolated galaxy or the best-resolved zoom-in cosmological simulations\footnote{We do not test at the resolution typical of very large cosmological simulations, since they typically cannot resolve star formation at the scale of 11 SN clusters, the only size for which we have 3D simulations.}) and high resolution ($\Delta r_0 = 0.6$ pc, sufficient to obtain a converged terminal momentum without an explicit subgrid recipe). While it would be ideal for each subgrid model to perfectly match the final momentum of a converged high-resolution 3D simulation, such a simulation is unfortunately not available for clustered SNe. In 1D, \citet{2017MNRAS.465.2471G} show that the converged result for the total momentum enhancement due to clustering is a factor of $\approx 10$ increase over simply adding up the momentum of single SNe (though this can be reduced if one includes an explicit model for turbulent conduction -- see \citealt{2019arXiv190209547E}), but \citet{2019MNRAS.483.3647G} show that 3D simulations remain unconverged even at $\approx 1$ pc resolution. However, the 3D result does provide a lower limit to the amount by which clustering enhances the terminal momenta of SNRs: by extrapolating from the simulation resolution to the Field length, they show that the expanded enhancement in terminal momentum per SN is at least a factor of $2-3$. This provides our primary criterion for the success of a subgrid model: when applied to a low resolution 11 SN cluster, does it result in a momentum efficiency \emph{at least} 2 times greater than the fiducial isolated SN momentum efficiency, thus reproducing the lower limit implied by the 3D simulations?

Finally, we needed to choose specific subgrid models to test. For this work we chose a sample of 3 commonly-used approaches: ``delayed cooling'' (specifically mimicking the ``blastwave'' feedback method available in the \texttt{GASOLINE-2} code; \citealt{2006MNRAS.373.1074S, 2017MNRAS.471.2357W}),
``momentum-energy injection'' (specifically mimicking the implementation used by the \texttt{FIRE-2} simulations; 
\citealt{2018MNRAS.480..800H}) and finally ``simultaneous energy injection'' (specifically mimicking the implementation used by the \texttt{EAGLE} simulations; \citealt{2012MNRAS.426..140D, 2015MNRAS.446..521S,2015MNRAS.450.1937C}). This is not meant to be an exhaustive list, but covers some of the most common approaches which could be tested by our code\footnote{For example, we cannot meaningfully test the subgrid model used in the \texttt{IllustrisTNG} simulations in a homogeneous ISM, as that model prescribes ``wind'' particles that travel from regions of high density and only inject their energy into the ISM in regions of low density \citep{2013MNRAS.436.3031V, 2018MNRAS.473.4077P}. Our homogeneous ISM has no regions of low density, so these wind particles would never re-couple with the ISM.}.

In \autoref{section:models} we introduce the ideas behind each of these subgrid models at a high level. 
In \autoref{section:methods} we discuss the numerical methods used in our 1D simulations, and then the implementation details needed for each subgrid model. 
In \autoref{section:results} we show the results of each simulation and briefly comment on the differences. 
We put these results in context in \autoref{section:discussion} and then conclude in \autoref{section:conclusions}.

%%%%%%%%%%%%%%%%%%%%%%%%%%%%%%%%%%%%%%%%%%%%%%

\section{Overview of Feedback Models}
\label{section:models}

Before getting into the implementation of each subgrid model, we give a conceptual overview of the physical motivation of each. In \autoref{section:methods} we cover these models again at a lower level, specifying the implementation details of each step.

\subsection{Delayed cooling feedback}
\label{section:overview:delayedcooling}

Since the underlying problem is that at low resolutions SNRs cool too quickly, before they are able to accelerate enough mass, one early approach was to simply ``turn off'' radiative cooling, allowing the SNR to develop until the correct radiative cooling timescale, and then turn cooling back on (at which point most of the energy is likely rapidly radiated away). Since this delays the cooling, we will refer to this as the ``delayed cooling'' model.

To focus on a specific example, we will mimic the model proposed by \citet{2006MNRAS.373.1074S} (although this idea dates back at least to \citealt{1997PhDT........19G}). These \citeauthor{1997PhDT........19G}-style models \citep[e.g.,][]{2000ApJ...545..728T} sought to be simpler than existing models that assumed unresolved phases within each resolution element in order to prevent overcooling, but they themselves introduced a number of free parameters which were not well explored until the systematic study by \citet{2006MNRAS.373.1074S}. \citeauthor{2006MNRAS.373.1074S} used observational constraints (primarily the slope, normalisation and cutoff of the Kennicutt-Schmidt star formation law \citep{Kennicutt98a}, and the rate and steadiness of star formation in the Milky Way over the past 1 Gyr) to optimise the free parameters of a delayed cooling model.

\citeauthor{2006MNRAS.373.1074S} found that the blast energy, $E_\mathrm{SN}$, is one of the hardest parameters to constrain. Even if a SN releases $10^{51}$ erg of energy in its ejecta, some of that energy should be radiated away (rather than becoming hydrodynamically coupled to the ISM), but since radiative cooling has been turned off in this model the blast energy likely needs to be decreased to compensate. \citeauthor{2006MNRAS.373.1074S} initially suggested $E_\mathrm{SN} = 10^{50}$ erg (i.e., 10\% of the ejecta energy), motivated by very high resolution simulations of isolated SNe \citep{1998ApJ...500...95T} and found that simulations using that value produce results consistent with the Milky Way hot gas fraction and a radial velocity dispersion comparable to observed quiescent spiral galaxies, leading them to recommend a default value of $E_\mathrm{SN} = 10^{50}$ erg. However, given the difficulty in constraining this parameter, we will test multiple values in this study.

\subsection{Momentum-energy feedback}

\label{section:overview:momentum_energy}

A second class a models takes a different tack: rather than focusing on intermediary steps (like mimicking the instantaneous cooling rate), we could instead focus on directly prescribing the key final results such as radial momentum. The radial momentum of a single SN expanding into a cold medium asymptotes to a value that is a function of the SN energy and the density and metallicity of the ambient medium \citep{1988ApJ...334..252C}, and this momentum is expected to be a key driver of small scale limits on star formation (e.g., by driving turbulence that maintains the scale height of a galactic disc; \citealt{2011ApJ...731...41O,2013MNRAS.433.1970F}), as well as large scale limits on star formation (e.g. galactic winds removing gas from galaxies; \citealt{2005ApJ...618..569M,2012MNRAS.421.3522H,2013MNRAS.432..455D,2013MNRAS.429.1922C,2016MNRAS.455..334T}).

The specific model we will mimic is from the \texttt{FIRE-2} methods (described in greatest depth by \citealt{2018MNRAS.477.1578H} and with more context by \citealt{2018MNRAS.480..800H}), although numerous earlier authors adopted very similar approaches (e.g., \citealt{2011ApJ...743...25K, 2014ApJ...788..121K, Kimm15a, 2015ApJ...809...69S,  2016ApJ...827...28G}).  \citet{2018MNRAS.477.1578H} set out to create and test a model that would work anywhere between moderately low and arbitrarily high resolutions, without requiring tunable parameters\footnote{Although the momentum-energy method we consider has no tunable parameters, it does require adopting a terminal momentum model. All of these models are simplifications that try to capture the primary dependencies of the different inputs tested (e.g., ISM density and metallicity), while assuming no untested factors play a significant role in determining asymptotic momentum.} (allowing them to make testable predictions of observables like galaxy mass, and star formation history whereas the delayed cooling model of \citealt{2006MNRAS.373.1074S} had to be forced to match these values by design).  In order to do this, they employed a hybrid method: at high resolution it is primarily a direct energy injection method (which requires very few assumptions), but at low resolution it transitions to a momentum injection method when needed.

The \texttt{FIRE-2} approach achieves this by first directly adding the ejecta to a neighbourhood around the SN location, calculating if that region has sufficient resolution to resolve the SNR evolution, and if not, adjusting the injected quantities (i.e. increasing the injected momentum and decreasing the injected energy), to approximate the late-time state of the SNR. When the expected cooling radius of a SN is unresolved, which is the case in almost all cosmological or galaxy-scale simulations focusing on spiral galaxies,\footnote{Though not in dwarfs, where lower densities yield larger cooling radii, and which can be simulated at much higher resolution due to their smaller overall size -- e.g., \citet{Forbes16a, 2018arXiv181202749W}.} this recipe essentially reduces to injecting a fixed amount of radial momentum per SN. In practice, this model will always apply ``energy-dominated'' feedback in the high resolution simulations presented in this paper, and will always start ``momentum-dominated'' in our low resolution simulations but by about the fourth SN even our low resolution 11 SN cluster will start receiving ``energy-dominated'' feedback as the central density decreases and the expected cooling radius increases.

\subsection{Simultaneous energy injection}
\label{section:overview:simultaneous}

The final class of models we will consider are those that try to address the mismatch between the mass that initially gains the energy in reality (i.e., the ejecta mass) and the mass that gains the energy in simulations (typically the kernel mass, which is much greater than the ejecta mass). For a fixed kernel mass, the first way to address a fixed amount of energy being spread too thinly between too much mass is to release more energy simultaneously into the fixed kernel, for example by injecting all SNe from a given cluster simultaneously.  An additional technique is to artificially limit the mass that receives energy to be smaller than the typical kernel mass. We will specifically mimic the approach proposed by \citet{2012MNRAS.426..140D} and used in the \texttt{EAGLE} simulations \citep{2015MNRAS.446..521S,2015MNRAS.450.1937C} which does both: simultaneously injects all the SN energy from a star cluster\footnote{\citeauthor{2012MNRAS.426..140D}  note that their model can be extended to stochastically release energy across \emph{multiple} timesteps rather than simultaneously, but since their work focuses on the simultaneous case, we classify it as a ``simultaneous'' model.}, and adding a stochastic element that only adds energy to a fraction of the kernel but in doing so \emph{guarantees} that any elements receiving energy are heated beyond the peak of the cooling curve.

The key parameter in the \texttt{EAGLE} approach is $\Delta \epsilon$, the increase in specific thermal energy by a cell receiving SN energy (mostly equivalent to a desired change in temperature, $\Delta T$). By making this a prescribed parameter, they can ensure that any resolution element that receives energy is sufficiently hot ($\gtrsim 10^{7.5} $ K) that the resolution element is radiatively inefficient. This comes at a cost; in order to inject the right amount of energy (on average), this sets a limit on the amount of mass (on average) of the resolution elements that can receive the energy. This is especially difficult at low resolution (for fixed cluster mass and SN energy), where we cannot usually select a set of resolution elements that partitions the injection kernel into the correct amount of mass that does and does not receive energy without splitting or merging resolution elements. To solve this, \citet{2012MNRAS.426..140D} suggest a stochastic approach.

Within an injection kernel, each cell has a probability of receiving energy, $p$; this probability depends on the mass within the kernel and the total blast energy, but does not vary between cells. For the $i^\mathrm{th}$ cell within the kernel (containing a cell mass $m_i$ after receiving its share of the SN ejecta mass) a thermal energy $m_i \Delta \epsilon$ is added with probability $p$. By prescribing a sufficiently high $\Delta \epsilon$, we can ensure that \emph{on average} we inject the correct energy, even if a particular cluster injects more or less than the desired amount. They recommend choosing a value of $\Delta \epsilon$ corresponding to $\Delta T = 10^{7.5}$ K, but note that for fixed total SN energy, higher resolutions will require higher values of $\Delta \epsilon$ (or else the model would calculate a value $p>1$ which leads to \emph{always} injecting too little energy). In Appendix~\ref{section:convergence} we discuss how resolution, total blast energy and the choice of $\Delta \epsilon$ affect the variance of the injected energy, which can guide the choice of $\Delta \epsilon$.

This shows that although the model is generally free of tunable parameters, a derivation from first-principles can only recommend a rough lower limit for $\Delta \epsilon$ rather than a specific value. In practice, they inform their choice of $\Delta \epsilon$ through use of a suite of galactic simulations, comparing observables such as morphology, phase distribution, star formation rate and large scale wind properties, meaning results from these observables in similar simulations using this method cannot be treated as original predictions.

%%%%%%%%%%%%%%%%%%%%%%%%%%%%%%%%%%%%%%%%%%%%%%

\section{Numerical Methods}
\label{section:methods}

% \subsection{1D simulations}
% \label{section:1D}

All the simulations we present here use the \texttt{clustered\_SNe} code described by \citet{2017MNRAS.465.2471G}, with a few modifications. We will briefly describe the general framework of the code and then clearly state the modifications, before moving onto the specific implementation details of each model added for this paper.

The \texttt{clustered\_SNe} code hydrodynamically evolves blasts in a 1D (spherically symmetric) environment, with the assumption that all SNe occur at the same location $r=0$. The ISM is assumed to have an initial spatially constant density (we use $\rho = 1.33 m_\mathrm{H}$ cm$^{-3}$ for all the simulations presented here) and metallicity ($Z = 0.02$ for all simulations here); we only track total metallicity, not any specific species.  The inner boundary is a zero-flux, zero-velocity boundary; the outer boundary condition does not matter since we choose a large enough domain so that the SNR does not reach the outer boundary.  This ISM mass and metallicity is placed within moving mesh cells, with an initial spacing of $\Delta r_0$ that depends on the specific simulation and no initial velocity. Since the spherical shells initially have equal thickness, they do not have equal initial volume or mass.

For a given cluster mass, the \texttt{clustered\_SNe} code uses the \texttt{SLUG} code \citep{2012ApJ...745..145D,2014MNRAS.444.3275D,2015MNRAS.452.1447K} to directly sample a \citet{Kroupa04012002} IMF of stars, and then explodes any stars with an initial mass greater than $8 M_\odot$ after a mass-dependent lifetime predicted by the Geneva stellar evolution models \citep{2012A&A...537A.146E}. Explosion mass and metal yields follow the results of \citet{2007PhR...442..269W}, while each explosion is assumed to yield a constant $E_\mathrm{blast} = 10^{51}$ erg of energy (unless otherwise specified by a feedback model).

Each SN is injected into the innermost cell (unless otherwise specified by a feedback model), and then the cells are evolved using an approximately Lagrangian HLLC solver (\citealt{HLLC}, with the specific implementation by \citealt{2016ApJ...821...76D}) 
solving for hydrodynamics but not other physics such as gravity,\footnote{By not including gravity we match our choice in our previous 3D simulations \citep{2019MNRAS.483.3647G}, but we note that in our previous 1D simulations \citep{2017MNRAS.465.2471G} we found that adding self-gravity decreased the terminal momentum by about a factor of 2 for this 11 SNe cluster.}
physical conduction or magnetohydrodynamics. Since cell boundaries are allowed to evolve, and arbitrarily thin cells would lead to arbitrarily small time steps and large computational costs, cells are merged with a neighbour when they become 10 times thinner than the average cell thickness (and cells are likewise split when 2 times thicker than the average cell thickness). Since the solver is only approximately Lagrangian, there is generally a small, non-zero mass flux between cells, but changes in cell mass should be minor outside of cell merge/split events and SN injection events. 
Optically-thin, metallicity-dependent radiative cooling is included using the \texttt{GRACKLE} cooling library \citep{2017MNRAS.466.2217S} assuming equilibrium chemistry and a \citet{2012ApJ...746..125H} extragalactic UV background but neglecting galactic heating sources.

We also made some modifications since the version described by \citet{2017MNRAS.465.2471G}. The biggest is that we changed the initial ISM temperature so that it matches the equilibrium temperature of the initial density and metallicity. For $\rho = 1.33 m_\mathrm{H}$ cm$^{-3}$, $Z = 0.02$, and a fixed $\gamma = 5/3$, this corresponds to a specific internal energy of $3.50 \times 10^{10}$ erg g$^{-1}$ (about 340 K as calculated by \texttt{GRACKLE}, where radiative cooling balances extragalactic UV heating while ignoring galactic heating sources). Although this choice has no significant effect on the simulation outcome, which is the same as long as the ambient temperature is much smaller than the temperature of the hot gas in the SNR interior, starting the gas in thermal equilibrium simplifies the analysis.  The second change is that, unlike in \citet{2017MNRAS.465.2471G}, we do not include pre-SN stellar winds; now the only mass that is injected is from the SN itself. We disable winds because they are generally not included in the feedback prescriptions we are testing.

For each feedback model we do simulations with 1 SN and 11 SNe, ensuring that the same SNe properties are drawn for all 1 and 11 SN simulations respectively. The 1 SN simulations are run until a cluster age of 20 Myr (roughly 10 Myr after the only SN); the 11 SN simulations are run until a cluster age of 100 Myr (roughly 97 Myr after the first SN). For each model and cluster size we also carry out both a high resolution run ($\Delta r_0 = 0.3$ for 1 SN;  $\Delta r_0 = 0.6$ for 11 SNe to match the reference run in \citealt{2017MNRAS.465.2471G}) and a low resolution run ($\Delta r_0 =$ 20 pc). The particular resolution of our low resolution run is chosen so that the region in the inner ``ghost'' cell (the innermost 20 pc, which we do not hydrodynamically evolve) contains $\sim 10^3 M_\odot$ of material, which would be expected to produce $\sim 11$ SNe if it were completely converted to stars.  This mass and spatial resolution is also comparable to the typical highest values achieved in modern zoom-in cosmological simulations of spiral galaxies. 
In Appendix~\ref{section:3SNe} we also show the results for a 3 SN cluster, but will not focus on it in the main body of this paper as we have no 3D 3 SN simulation to test the results against. 

\subsection{Delayed cooling implementation}
\label{section:methods:delayedcooling}

As described in \autoref{section:overview:delayedcooling}, the key idea behind the delayed cooling model is that there should be a spatial scale around the location of a SN into which the SNR can expand before losing a significant amount of energy to radiative cooling; associated with this expansion should also be a characteristic time scale.  The approach of the delayed cooling model is to temporarily disable radiative cooling for any resolution elements initially within this spatial scale, $R_\mathrm{E}$ for an appropriate time scale $t_\mathrm{E}$, explicitly delaying cooling.

These key scales are given as analytic expressions by \citet[their Equations 9 and 10]{2006MNRAS.373.1074S} that depend on the local ISM density, $\rho_0$, and pressure, $P_0$ as well as the blast energy\footnote{When multiple SNe occur within the same timestep, the respective blast energies are typically combined for delayed cooling models but this does not happen for the cluster parameters and resolutions we consider here.}, $E_\mathrm{blast}$:
\begin{align}
    R_\mathrm{E}(\rho_0, P_0) =& 10^{1.74} \left( \frac{E_\mathrm{blast}}{10^{51} \mathrm{erg}} \right)^{0.32} \left(\frac{\rho_0}{ \mu m_\mathrm{H} \text{ cm}^{-3}}\right)^{-0.16} 
    \nonumber\\ &  \times
    \left(\frac{P_{0}}{ 10^4 k_\mathrm{B} \text{ K}    \text{ cm}^{-3} }\right)^{-0.20} \text{ pc}
    \label{eq:RE}
\end{align} 
\begin{align}
    t_\mathrm{E}(\rho_0, P_0) =& 10^{5.92} \left(\frac{E_\mathrm{blast}}{10^{51} \mathrm{erg}} \right)^{0.31} \left(\frac{\rho_0}{ \mu m_\mathrm{H} \text{ cm}^{-3}}\right)^{0.27}
    \nonumber \\ & \times
    \left(\frac{P_{0}}{ 10^4 k_\mathrm{B} \text{ K}    \text{ cm}^{-3} }\right)^{-0.64} \text{ yr}
    \label{eq:tE}
\end{align}
where $\mu$ is the mean molecular weight, $m_\mathrm{H}$ is the mass of the hydrogen atom, $k_\mathrm{B}$ is the Boltzmann constant. (Note: \citet{2006MNRAS.373.1074S} end up adopting a slightly different, slightly longer timescale for their final model, but in their Section 5.3.1 they ultimately conclude it should not make a significant difference.)

It is easy to evaluate these expressions for the first SN when there is a single, clear value for $\rho_0$ and $P_0$ due to our initially-homogeneous ISM; for subsequent SNe it becomes more ambiguous due to the non-uniform bubble that has formed.  To handle this, we solve for $R_\mathrm{E}$ iteratively, starting from the centre of the simulation and stepping outwards until the volume-weighted average density and pressure result in an $R_\mathrm{E}$ that matched the current radius.  Fortunately, preliminary tests (using the results from an 11 SN reference run) found that there should typically be a single, unique $R_\mathrm{E}$ for each SN; in practice we take the first valid $R_\mathrm{E}$ (i.e.  $R_\mathrm{E}(\rho_0(<r), P_0(<r)) < r$). Within this radius, all cells have their cooling disabled for a time $t_\mathrm{E}$, even if they later move beyond this radius. Cells that have already had their cooling disabled from a previous SN have their cooling turned off for the longer of: the current SN's $t_\mathrm{E}$ and the remaining duration from the previous SNe's shutoff periods ($t_{\mathrm{E}, i} - t_{\mathrm{SN}, i}$).

Since the cooling rate of a SNR does not truly go to 0 at high resolution, \citet{2006MNRAS.373.1074S} leave $E_\mathrm{blast}$ as a free parameter that can be decreased to compensate, suggesting a typical value of $E_\mathrm{blast} = 10^{50}$ based off their initial experiments. For each cluster we simulate, we will run two variants in order to explore the effect of this free parameter: one using  $E_\mathrm{blast} = 10^{50}$ and another using $E_\mathrm{blast} = 10^{51}$.

In addition to $R_\mathrm{E}$ and $t_\mathrm{E}$, we must compute where to deposit the SN energy and mass. Following \citet{2006MNRAS.373.1074S}, we assume a fixed kernel mass ($M_\mathrm{kernel} = 3 \times 10^5 M_\odot$), and solve for the corresponding radius, $R_\mathrm{kernel}$, that encloses that mass. Within this radius, we inject mass and energy using a Gaussian kernel with 1D dispersion $\sigma = 0.1 R_\mathrm{kernel}$, weighted by the cell masses and truncated at $R_\mathrm{kernel}$.  This kernel mass was chosen so that $R_\mathrm{kernel}$ is always larger than $R_\mathrm{E}$; this means some energy will be injected outside the cooling-disabled region and will be lost rapidly, but the amount so affected is minimal due to the sharp drop off in the Gaussian profile.

Finally, we point out that it is important that when the first SN occurs, the ISM is near its equilibrium temperature rather than significantly above it (which was originally the default of the \texttt{clustered\_SNe} code; \citealt{2017MNRAS.465.2471G}). If the ISM is far from equilibrium with a short cooling time, then an artificial discontinuity would rapidly develop near $R_\mathrm{E}$ after the first SN. Within $R_\mathrm{E}$ the gas would stay hot and over-pressured, while just beyond $R_\mathrm{E}$ the gas would cool and drop in pressure, leading to an outward-propagating shock.

\subsection{Momentum-energy feedback implementation}
\label{section:methods:momentum_energy}

Momentum-energy models are characterised by a common thread: at high resolution their key active component is the injected energy, while at low resolution the active component is injected radial momentum, with a continuous transition between these regimes. For this work, we base our implementation on the \texttt{FIRE-2} algorithm \citep{2018MNRAS.477.1578H}, but make a few necessary alterations to match the different geometry of our cells. Unlike \citet{2018MNRAS.477.1578H} we neglect stellar winds in this work.

First, we define the total SN mass, momentum, energy and metal mass yields.  The SN-ejected mass and metallicity will be consistent with our reference simulations, along with the times at which these SNe occur; this is in contrast to the SN mass yields and delay times suggested by \citet{2018MNRAS.477.1578H} although we expect this makes relatively little difference.  Next, we keep the SN blast energy at the fiducial $E_\mathrm{blast}=10^{51}$ erg.  The biggest difference between this model and our  reference simulations is that this model also injects momentum. At high resolution, this momentum is determined by assuming the blast energy is fully kinetic:
\begin{equation}
    p_\mathrm{ejecta} = \sqrt{2 m_\mathrm{ejecta} E_\mathrm{ejecta}}
\end{equation}
but this momentum will be increased at lower resolution, depending on the properties of the cells into which the blast is injected. As a reminder, although we say the ejecta might be ``fully kinetic'', we only explicitly track metal mass, total mass, momentum and total energy---not kinetic and thermal energy separately. So even if the ejecta energy is ``fully kinetic'', the ejecta is assumed to inelastically collide with the cell mass, causing a change in both kinetic and thermal energy. By only tracking total energy, we do not need to directly prescribe how this collision occurs; we can simply solve for the resulting kinetic and thermal energy using the resulting cell mass, momentum and total energy.

The injection kernel is probably the most significant departure from the algorithm described by \citet{2018MNRAS.477.1578H} owing to the different geometry of our simulation (we use rigidly structured 1D, nested shells, whereas the \texttt{FIRE-2} simulations use unstructured 3D moving particles). We identify the injection kernel to comprise the innermost $N_\mathrm{ngb} =3$ cells ($\approx 32^{1/3}$, as opposed to their 32 nearest neighbours). Next, we weight all cells equally, $w_i = N_\mathrm{ngb}^{-1}$ (using index $i$ for the $i$th cell); this is in contrast to their solid angle-based weighting \citep[Eq. 2]{2018MNRAS.477.1578H} which would only ever inject into 1 cell given our enforced spherical symmetry. Since we used a fixed $N_\mathrm{ngb}$ independent of resolution, the weights are then equal and also independent of resolution.

We now can specify the amount of mass, metals, momentum and energy added to each individual cell. Mass and metals are easy; they follow the weights: 
\begin{equation}
    \Delta m_i = w_i m_\mathrm{ejecta}
\end{equation}
and
\begin{equation}
    \Delta m_{Z, i} = w_i m_{Z, \mathrm{ejecta}} .
\end{equation}
The injected momentum is slightly more complicated.  As mentioned above, we start with a base amount of total momentum, $p_\mathrm{ejecta}$, but as the resolution decreases the amount of momentum is increased to mimic the SNR evolution below the resolved scales. At arbitrarily low resolution, all stages of the SNR evolution will be unresolved, so the momentum should approach the expected terminal momentum. This expected terminal momentum is calculated for each cell based on the cell's gas density and metallicity:
\begin{align}
	\frac{p_{\mathrm{t},i}}{ M_\odot \text{ km s}^{-1}} = & 4.8 \times 10^5 \left( \frac{E_\mathrm{blast}}{10^{51} \mathrm{erg}} \right)^{13/14} 
	\nonumber \\ & \times
	\left( \frac{\rho_i}{\mu_i m_\mathrm{H} \text{ cm}^{-3}} \right)^{-1/7} f(Z_i)^{3/2}
	\label{eq:p_t_i}
\end{align}
where
\begin{equation}
	f(Z) = \begin{cases}
	2 & Z / Z_\odot < 0.01
	\\
	(Z / Z_\odot)^{-0.14} & \mathrm{otherwise}
	\end{cases}
\end{equation}
and $Z_\odot = 0.02$.
The two extremes---arbitrarily high resolution for which $\Delta p_i = w_i p_\mathrm{ejecta}$ and arbitrarily low resolution for which $\Delta p = w_i p_{\mathrm{t},i}$---are tied together by:
\begin{equation}
    \Delta p_i = w_i p_\mathrm{ejecta} \min \left(\sqrt{1 + \frac{m_i}{\Delta m_i}}, \frac{p_{\mathrm{t}, i}}{p_\mathrm{ejecta}} \right) 
\end{equation}

The injected total energy similarly starts with a high resolution proposal: $\Delta E_i = w_i E_\mathrm{blast}$ which is then corrected at low resolutions.  The motivation for this correction is to avoid adding energy at unphysical large distances from the SN. While a SNR will not be able to directly add energy beyond the cooling radius of the SNR, at very low resolution, some of the $N_\mathrm{ngb}$ neighbours in which we deposit energy might be beyond this cooling radius. Therefore, for each cell we first compute the expected SNR cooling radius:
\begin{equation}
    R_{\mathrm{cool}, i} = 28.4 \left(\frac{\rho_i}{\mu_i m_\mathrm{H} \text{ cm}^{-3}}\right)^{-3/7}  \left(\frac{E_\mathrm{blast}}{10^{51} \mathrm{erg}} \right)^{2/7} f(Z_i) \text{ pc}.
\end{equation}
If the cell's distance from the SN, $r_i$, is larger than $R_{\mathrm{cool}, i}$, we then calculate  how much the injection event will cause the cell's internal energy to change, $\Delta U_i$, and reduce the total energy injected so that the cell's change in internal energy is reduced by a factor  $(r_i / R_{\mathrm{cool},i})^{-6.5}$ while the cell's change in kinetic energy is unaffected.

\subsection{Simultaneous energy injection implementation}
\label{section:methods:simultaneous}

As introduced above in \autoref{section:overview:simultaneous}, the simultaneous injection model attempts to explicitly harness the clustering of SNe by forcing every SN from a cluster to occur simultaneously. This loss of time-resolution changes the dynamics, hopefully in a positive way. Rather than each SN failing to heat the nearby material past the peak of the cooling curve (and quickly overcooling as a result), injecting all SN energy at the same time makes it more likely that the affected material will be heated beyond the peak of the cooling curve. \citet{2012MNRAS.426..140D}  go one step further, introducing a stochastic component that guarantees material is heated past the peak of the cooling curve; it is this particular algorithm that we will try to follow as closely as possible.

The first part, defining the SN yields and delay time distribution, is easy. The yields are the same as in our reference model, in particular $E_\mathrm{blast} = 10^{51}$ erg. For the delay time distribution, if there are multiple SNe we modify the explosion times to all occur at $t=30$ Myr, matching \citet{2012MNRAS.426..140D}; if there is only 1 SN, we do not modify the explosion time (typically $t\approx 10$ Myr), since that corresponds to an arbitrary shift of when we define $t=0$ and does not affect the results in any way.

The injection kernel comprises the innermost $N_\mathrm{ngb} = 3$ cells\footnote{Conveniently, in our low resolution 11 SN simulations, our ghost cell mass matches the cluster mass, $m_\star$,  and the nearest 3 cells enclose $m_\mathrm{kernel} \approx 70 m_\mathrm{star}$, the closest we can get at this resolution to the value $m_\mathrm{kernel} = 48 m_\star$ used by \citet{2012MNRAS.426..140D}.}, into which mass and metals are injected deterministically, while energy is injected stochastically. The mass and metals are distributed evenly between each cell: 
\begin{equation}
    \Delta m_i = N_\mathrm{ngb}^{-1} m_\mathrm{ejecta}
\end{equation}
and
\begin{equation}
    \Delta m_{Z, i} = N_\mathrm{ngb}^{-1} m_{\mathrm{ejecta}, Z}.
\end{equation}

The stochastic energy injection within this kernel is more complicated. First, the mass within the kernel, $m_\mathrm{kernel}$ is computed, after adding the SN ejecta material. Then, given a value of $\Delta \epsilon$, the desired increase in specific thermal energy, we can calculate the probability of any cell within the kernel receiving energy:
\begin{equation}
    p = \frac{E_\mathrm{blast} N_\mathrm{SNe}}{\Delta \epsilon} \frac{1}{m_\mathrm{kernel}}
\end{equation}

At high resolution the restriction $p<1$ becomes problematic given the small kernel mass ($\sim 14 M_\odot$ for 1 SN and $130 M_\odot$ for 11 SNe, coming predominantly from the ejecta mass). So at high resolution, we adopt a value $\Delta \epsilon$ corresponding to a $\Delta T = 10^9$ K. 

At low resolution, the $p<1$ constraint is not a problem since the kernel masses are large: $\sim 70 \times 10^3 M_\odot$ for both our 1 SN and 11 SN clusters (the kernel masses are approximately identical for both setups because the mass is dominated by material already in the ISM before the SN(e), and the ISM mass within the kernel is only determined by $N_\mathrm{ngb}$ and the initial resolution and density which are all identical for the two clusters by construct). For 11 SNe, we can use $\Delta T = 10^{7.5}$ K as recommended by \citep{2012MNRAS.426..140D}, since in this case our kernel mass to cluster stellar mass ratio roughly matches the value they used.  For 1 SN at low resolution, the kernel mass is not as well matched. In order to have a reasonable probability of injecting \emph{any} energy we have to reduce $\Delta \epsilon$ to the value corresponding to $\Delta T = 3 \times 10^6$ K. (Even with a $\Delta T$ as low as $3 \times 10^6$ K, there is still an $\approx 90$\% chance that no cell receives energy for our 1 SN, low resolution simulations.) This value of $\Delta T$ is sufficiently high to result in a cooling time that is about an order of magnitude large than the sound crossing time, satisfying the condition suggested by \citet[Section 4]{2012MNRAS.426..140D}: that the cooling time of a resolution element must exceed its sound crossing time in order to to avoid artificial cooling.

Although in a galactic simulation we would select cells independently and stochastically with probability $p$, in this controlled test we will simply run all 8 possible realisations deterministically for a given number of SNe and resolution (enumerating the $2^3$ possibilities per setup of selecting or not selecting 3 cells---so a total of 32 simulations: 8 for each row in \autoref{tab:results}). Each of these realisations will be weighted with their respective Bernoulli probability: $p^N (1-p)^{3-N}$ for a total of $N$ cells being selected for energy injection.

%%%%%%%%%%%%%%%%%%%%%%%%%%%%%%%%%%%%%%%%%%%%%%

\section{Results of Feedback Models}
\label{section:results}

%!TEX root = feedback_models.tex

\begin{table*}
\caption{ Simulation results. ``Model'' refers to the injection model; $N_\mathrm{SNe}$ refers to the number of SNe from the cluster; $E_\mathrm{blast}$ denotes the energy added \emph{per SNe}; $\Delta r_0$ gives the initial spatial resolution; $t_\mathrm{end}$ indicates when we extract the final momentum and energy of the simulation relative to the time of cluster formation; $p_\mathrm{end}$ gives the momentum at that time; $E_\mathrm{kin}$ is the final total kinetic energy within the computational domain while $\Delta E_\mathrm{int}$ is the change in total internal energy within the computational domain relative to the moment before the first SN. For the simultaneous energy injection model, the final momentum and energy results are presented as the mean and standard deviation of our realisations (but it should be remembered that the distributions of these results are very non-gaussian; see Figures~\ref{fig:simultaneous:final:1} and \ref{fig:simultaneous:final:11}).
}
\label{tab:results}

\begin{tabular}{lrrrrrrrrrrc}
Model & $N_\mathrm{SNe}$ & $E_\mathrm{blast}$  & $\Delta r_0$ & $t_\mathrm{end}$ & $p_\mathrm{end} / N_\mathrm{SNe}$ &  $E_\mathrm{kin}$ & $\Delta E_\mathrm{int}$ \\
 & & (erg) & (pc) & (Myr)  & ($100 M_\odot$ km s$^{-1}$) & ($10^{49}$ erg) & ($10^{49}$ erg)  \\
\hline

\multirow{4}{*}{Reference} & 1 & $  10^{51} $  & 0.3 & 20  & 2763 & 0.620 & -0.071 \\
 & 1 & $  10^{51} $  & 20.0 & 20  & 265 & 0.008 & -2.856 \\
 & 11 & $  10^{51} $  & 0.6 & 100  & 34037 & 61.483 & 23.919 \\
 & 11 & $  10^{51} $  & 20.0 & 100  & 9153 & 11.677 & 3.354 \\

\hline

\multirow{8}{*}{Delayed cooling} & 1 & $  10^{50} $  & 0.3 & 20  & 3803 & 0.972 & -0.002 \\
 & 1 & $  10^{50} $  & 20.0 & 20  & 2139 & 0.320 & -1.252 \\
 & 1 & $  10^{51} $  & 0.3 & 20  & 37058 & 18.274 & -0.406 \\
 & 1 & $  10^{51} $  & 20.0 & 20  & 31609 & 12.833 & 0.159 \\
 & 11 & $  10^{50} $  & 0.6 & 100  & 3939 & 4.065 & 2.335 \\
 & 11 & $  10^{50} $  & 20.0 & 100  & 1805 & 1.458 & 0.585 \\
 & 11 & $  10^{51} $  & 0.6 & 100  & 34185 & 60.960 & 26.573 \\
 & 11 & $  10^{51} $  & 20.0 & 100  & 25169 & 35.023 & 315.989 \\

\hline

\multirow{4}{*}{Momentum-energy} & 1 & $  10^{51} $  & 0.3 & 20  & 2666 & 0.597 & -0.063 \\
 & 1 & $  10^{51} $  & 20.0 & 20  & 4047 & 0.873 & -0.275 \\
 & 11 & $  10^{51} $  & 0.6 & 100  & 33770 & 62.205 & 29.932 \\
 & 11 & $  10^{51} $  & 20.0 & 100  & 14010 & 18.326 & 5.409 \\

\hline

\multirow{4}{*}{Simultaneous} & 1 & $  10^{51} $  & 0.3 & 20  & $2781 \pm 1414$ & $0.663 \pm 0.371$ & $-0.075 \pm 0.059$ \\
 & 1 & $  10^{51} $  & 20.0 & 20  & $373 \pm 955$ & $0.076 \pm 0.240$ & $-1.818 \pm 0.653$ \\
 & 11 & $  10^{51} $  & 0.6 & 100  & $2070 \pm 870$ & $2.427 \pm 0.982$ & $0.527 \pm 0.265$ \\
 & 11 & $  10^{51} $  & 20.0 & 100  & $874 \pm 2700$ & $7.845 \pm 33.433$ & $-21.528 \pm 13.502$ \\

\end{tabular}
\end{table*}

In this section we give the results and discuss each model in turn. In the next section (\autoref{section:discussion}), we compare the results between models. A summary of the simulation results is given in \autoref{tab:results}.

\subsection{Direct injection results}
\label{section:results:direct}

\begin{figure}
\centering
\includegraphics[width=\columnwidth]{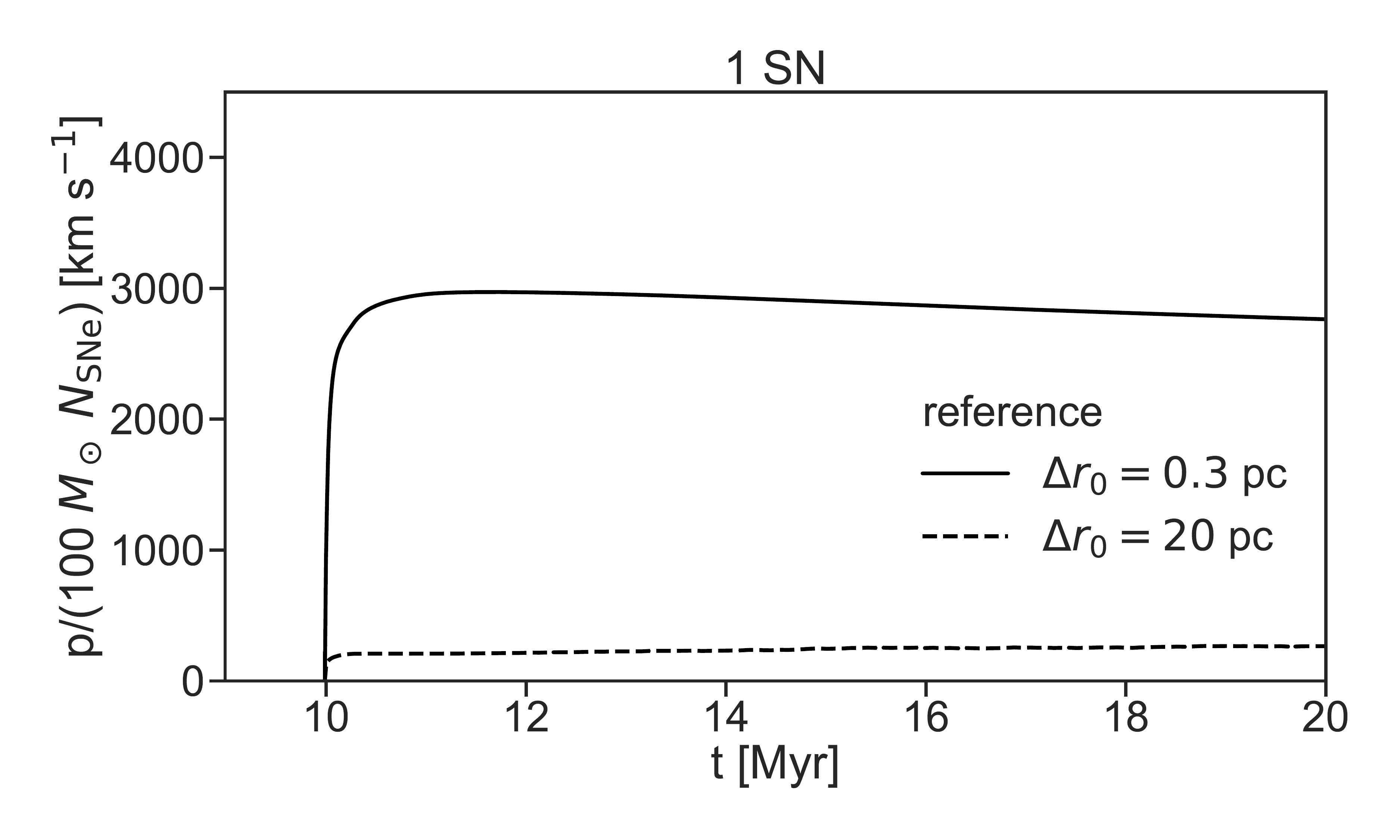}
\includegraphics[width=\columnwidth]{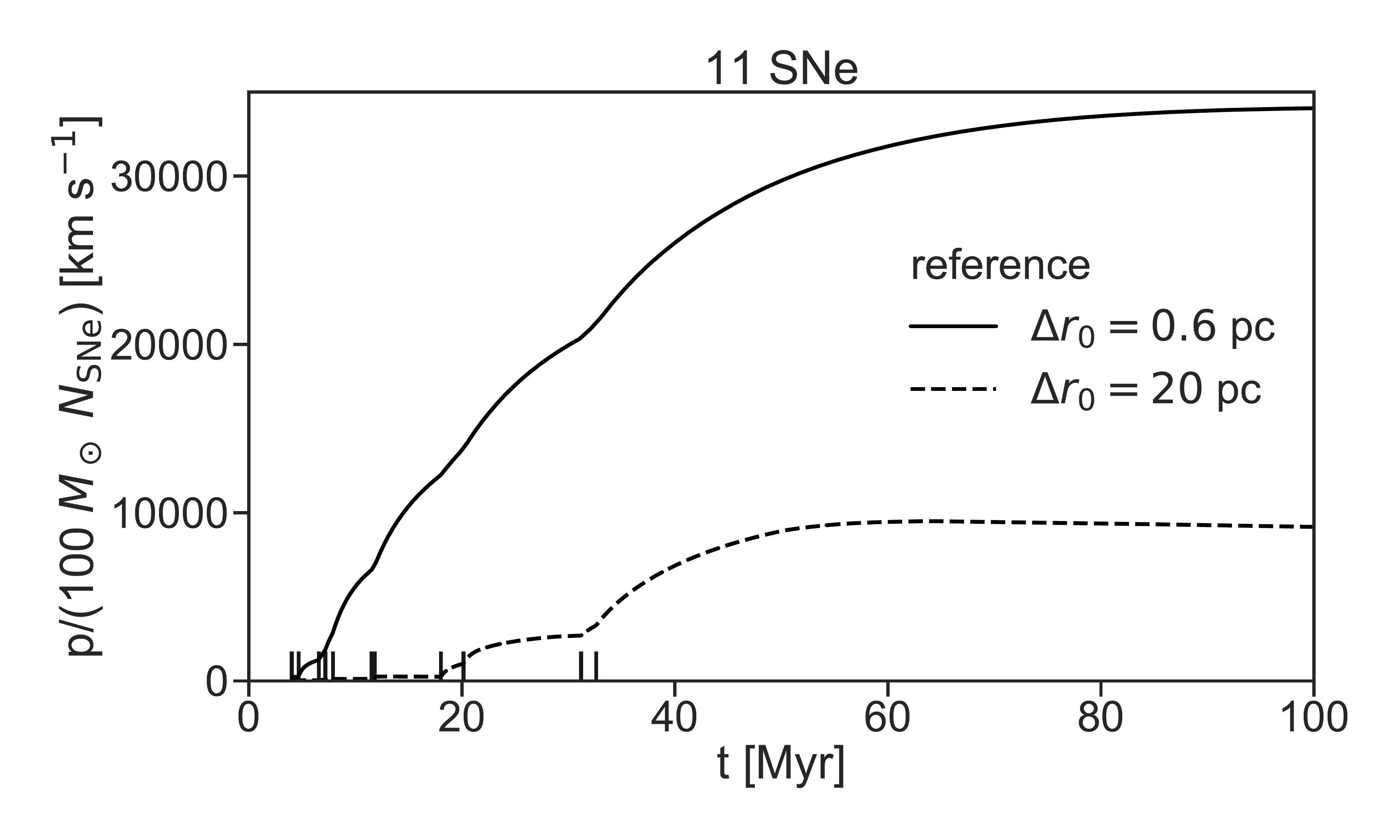}
\caption{
Comparison of the momentum as a function of time for our direct injection simulations, which we later use as the reference results. The top shows a comparison between different resolution 1 SN simulations; the bottom shows the different resolution 11 SN simulations. Note the difference in both horizontal and vertical scale between the two panels, both here and in subsequent similar plots.
}
\label{fig:reference:evolution}
\end{figure}

We start with the simplest simulations which directly inject thermal energy and no momentum to the innermost cell. The evolution of the momentum with respect to time for these 4 simulations is shown in  \autoref{fig:reference:evolution}. These will provide the ``reference'' results, against which we will compare the 3 competing models.

Looking first at the 1 SN results, we see the standard picture: at high resolution, we recover the standard terminal momentum ($\sim 3000$ $M_\odot$ km s$^{-1}$), but at low resolution, overcooling becomes a problem, leading to far too little momentum (in this case, an order of magnitude too little momentum).

This result is mirrored in the 11 SN simulations, but there it is slightly mitigated. As subsequent SNe occur, the density near the location of the blast drops, causing cooling to become less efficient and the resolution requirements to be loosened. This is not a perfect solution; it takes multiple SNe before a well-defined superbubble is inflated at low resolution, and even then each SN blast eventually propagates out to the dense shell where overcooling can occur. Still, the low resolution 11 SN run only contains 3-4 times less momentum than the high resolution simulation, whereas the 1 SN simulations exhibit a factor of 10 deficit in momentum when the resolution is worsened.

\subsection{Delayed cooling results}
\label{section:results:delayedcooling}

\begin{figure}
\centering
\includegraphics[width=\columnwidth]{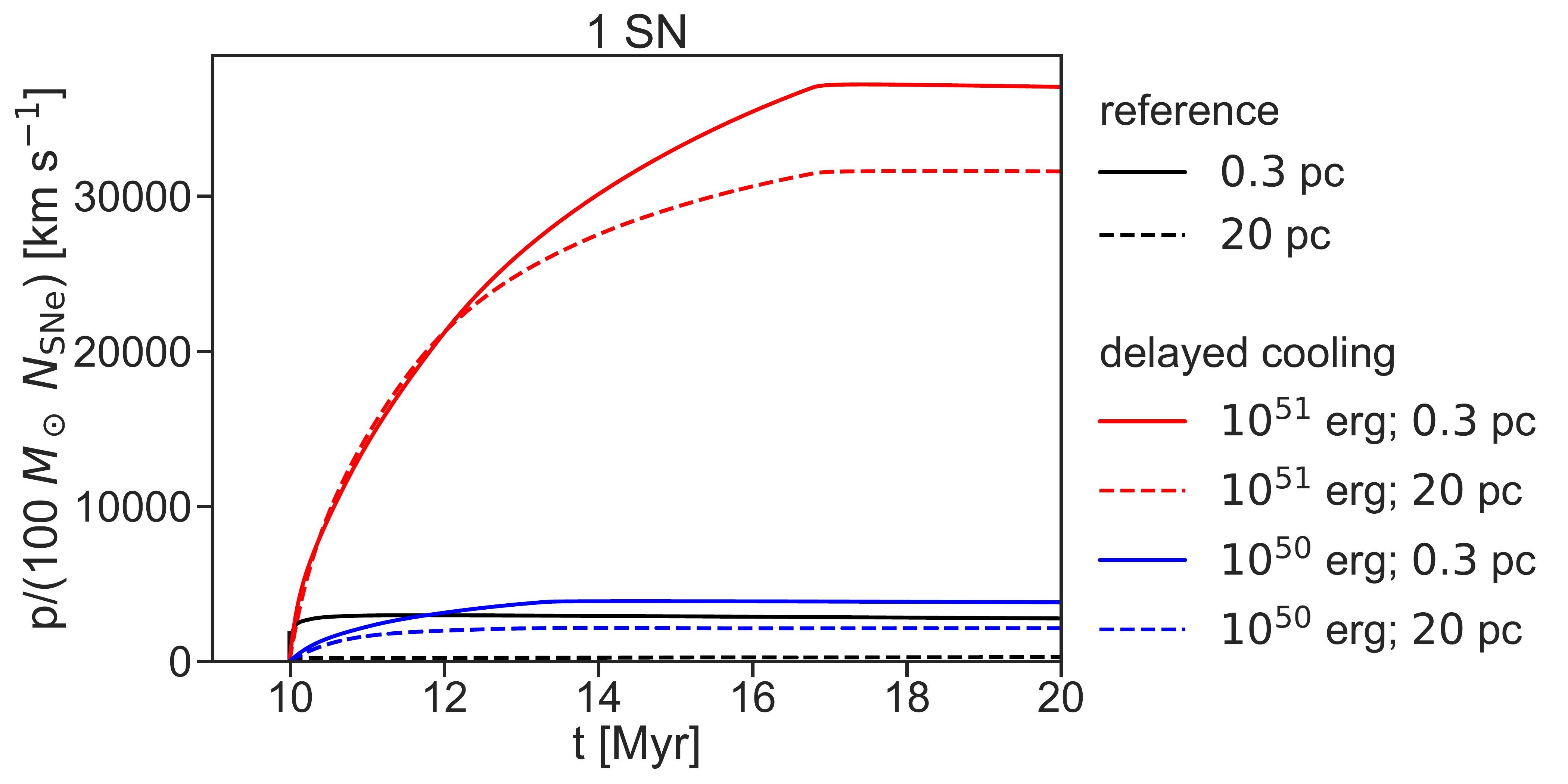}
\includegraphics[width=\columnwidth]{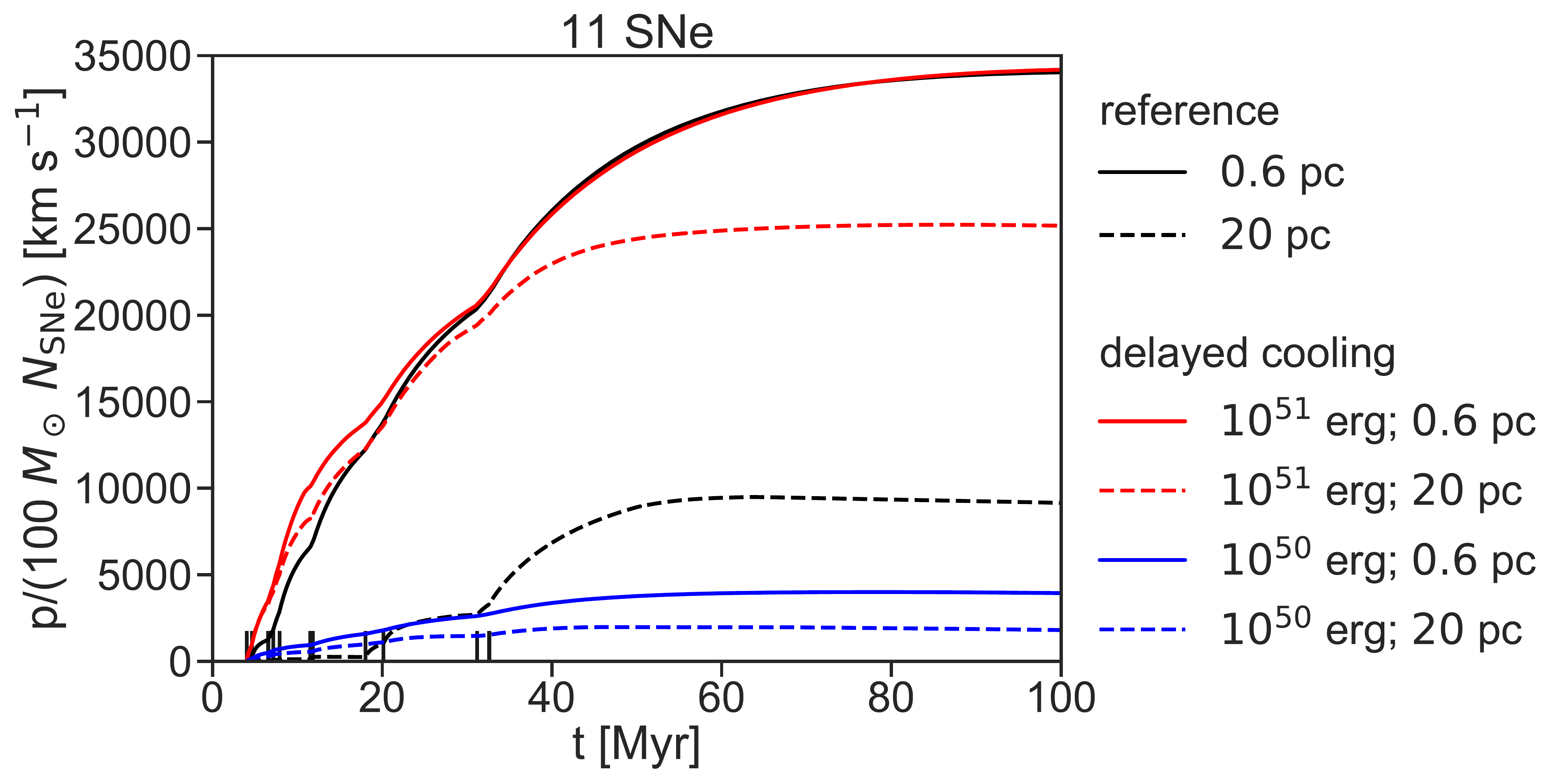}
\caption{
Same as \autoref{fig:reference:evolution}, except now also overplotting the delayed cooling simulations, with the different resolutions and $E_\mathrm{blast}$ parameters denoted in the legends. 
}
\label{fig:delayedcooling:evolution}
\end{figure}

Similar to the reference case, we ran simulations of 1 and 11 SNe at high and low resolution.  Since the blast energy is a free parameter within the \citet{2006MNRAS.373.1074S} algorithm, we carry out this experiment for two different blast energies: $E_\mathrm{blast} = 10^{50}$ erg and $10^{51}$ erg.  The resulting momentum evolution of each simulation can be seen in \autoref{fig:delayedcooling:evolution}.

Starting with the 1 SN simulations, we see the expected behaviour. When injecting the \emph{recommended} blast energy ($10^{50}$ erg), both the high resolution \emph{and} the low resolution simulations do a good job reproducing the terminal momentum of the high resolution reference simulation. This is what the method was designed to do. Conversely, when we inject a blast of $10^{51}$ erg and also shut off cooling, we observe a terminal momentum that is too large by a factor of $\sim 10$. This is unsurprising for the high resolution case; if we add the same amount of energy, but delay the onset of cooling, the result will be too much momentum. It is slightly more interesting that the low resolution simulation \emph{also} results in too much momentum. However, \citet{2006MNRAS.373.1074S} were aware of this potential problem, and this is what motivated them to recommend decreasing the blast energy.

The situation for the 11 SN simulations is very different.  In this case using $E_\mathrm{blast} = 10^{50}$ erg results in far \emph{too little} momentum---at high resolution it even does worse than our low resolution reference simulation (which used $E_\mathrm{blast} = 10^{51}$ erg).  This is because for the later SNe, the superbubble approaches near-adiabatic behaviour. Since the cooling radius is within the bubble where cooling is already relatively low, the cooling shut off switch in the delayed cooling prescription has little effect. However, the reduction in injection energy that \citet{2006MNRAS.373.1074S} recommend in order to fix the single-SN case then results in an under-powered superbubble and too little momentum even at high resolution.
(Note that even in the ``high'' resolution simulations by \citet{2006MNRAS.373.1074S}, a typical cluster mass is $10^4 M_\odot$, resulting in $\sim 100 $ SNe within 30 Myr, which means on average greater than 1 SN per timestep (they typically calculate star formation every 1 Myr). This means that rather than a series of SNe with relatively smaller cooling radii, SNe will be grouped into fewer but higher energy blasts, with larger cooling radii and longer cooling timescales (see their Figure 18). This might help explain why their recommended blast energy proved sufficient for their much lower resolution simulations, but we find that it is not powerful enough for our simulations.)

When we use stronger blasts ($E_\mathrm{blast} = 10^{51}$ erg), we find better results for the 11 SN case. For the first few SNe, the momentum starts too high (as expected from our 1 SN simulation results), but as the superbubble becomes more adiabatic, the delayed cooling approach starts to approach our direct injection behaviour. It is remarkable just how well the delayed cooling approach does with $E_\mathrm{blast} = 10^{51}$ erg for 11 SNe. Despite starting with more momentum at early times, and still shutting off cooling for each subsequent SN, the high resolution 11 SN simulation with $E_\mathrm{blast} = 10^{51}$ differs in terminal momentum from the high resolution reference simulation by less than 1\%. The lower resolution delayed cooling simulation ($E_\mathrm{blast} = 10^{51}$) does not do quite as well, but still is relatively close (differing by only about 25\%).

Thus this method is successful at being \emph{less resolution-dependent} than the reference, direct injection method. Whether this resolution-robust method produces accurate momenta is a more complicated question. We have found that a fixed $E_\mathrm{blast}$ is unable to handle both 1 SN and 11 SN clusters; an energy of $E_{\rm blast}=10^{50}$ erg as recommended by \citet{2006MNRAS.373.1074S} gives a good fit to the single SN case, but fails for 11 SNe, while injecting the full SN energy $E_{\rm blast}=10^{51}$ erg succeeds for 11 SNe but fails for 1. The fundamental reason for this is easy to understand: the mean amount of radiative loss per SN is not a single number, but instead depends on the SN environment, and in particular on whether a SN is going off inside an already low-density, hot cavity carved by a previous SN. Both the \textit{ad hoc} reduction from $E_{\rm blast} = 10^{51}$ erg to $10^{50}$ erg and the density- and pressure-dependence built into \autoref{eq:RE} and \autoref{eq:tE} for $R_\mathrm{E}$ and $t_\mathrm{E}$ attempt to capture the complex dependence of radiative loss on environment, but they do not do so with sufficient accuracy to reproduce the results of the high resolution simulation across a factor of 10 in cluster size. That said, this analysis suggests that it might be possible to find a  prescription for $E_\mathrm{blast}(N_\mathrm{SNe})$, or to choose a value of $\langle E_\mathrm{blast} \rangle$ averaged over the cluster mass function, that performs better than the current approximation of picking a single $E_{\rm blast}$. This would require a campaign of simulations similar to ours, to quantity the amount of radiative loss as a function of cluster size.

This is a clear indication, and a reminder, that these subgrid models can behave differently at different cluster sizes. This both means that it is useful to check how well each method is able to handle clustered SNe, but also provides a warning that our conclusions likely depend on the cluster sizes we test here (1 SN and 11 SNe). \citet{2019arXiv190300962D} identify as many as seven possible SN-driven superbubble regimes, of which we have tested only two.

\subsection{Momentum-energy feedback results}
\label{section:results:momentum_energy}

\begin{figure}
\centering
\includegraphics[width=\columnwidth]{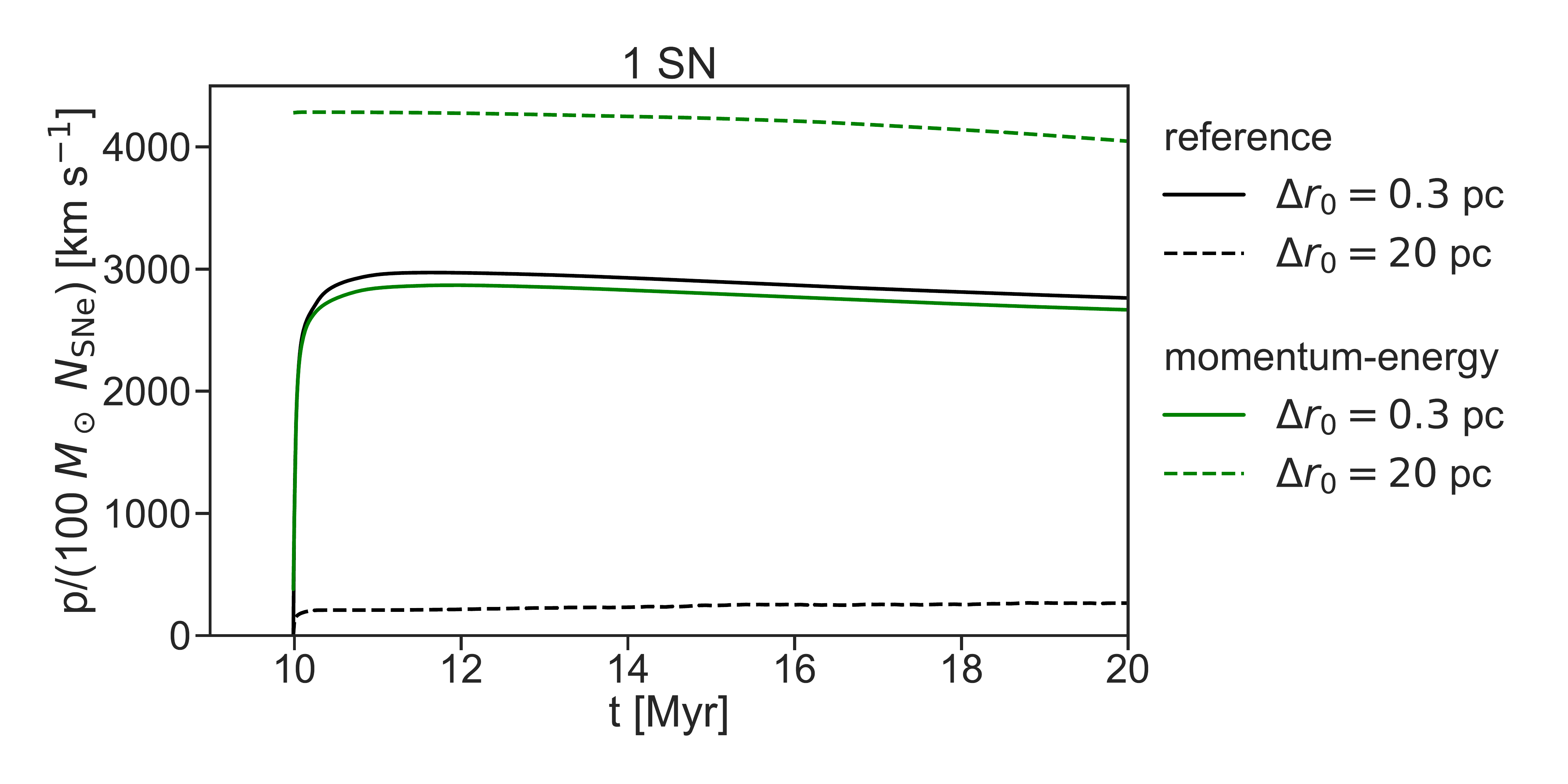}
\includegraphics[width=\columnwidth]{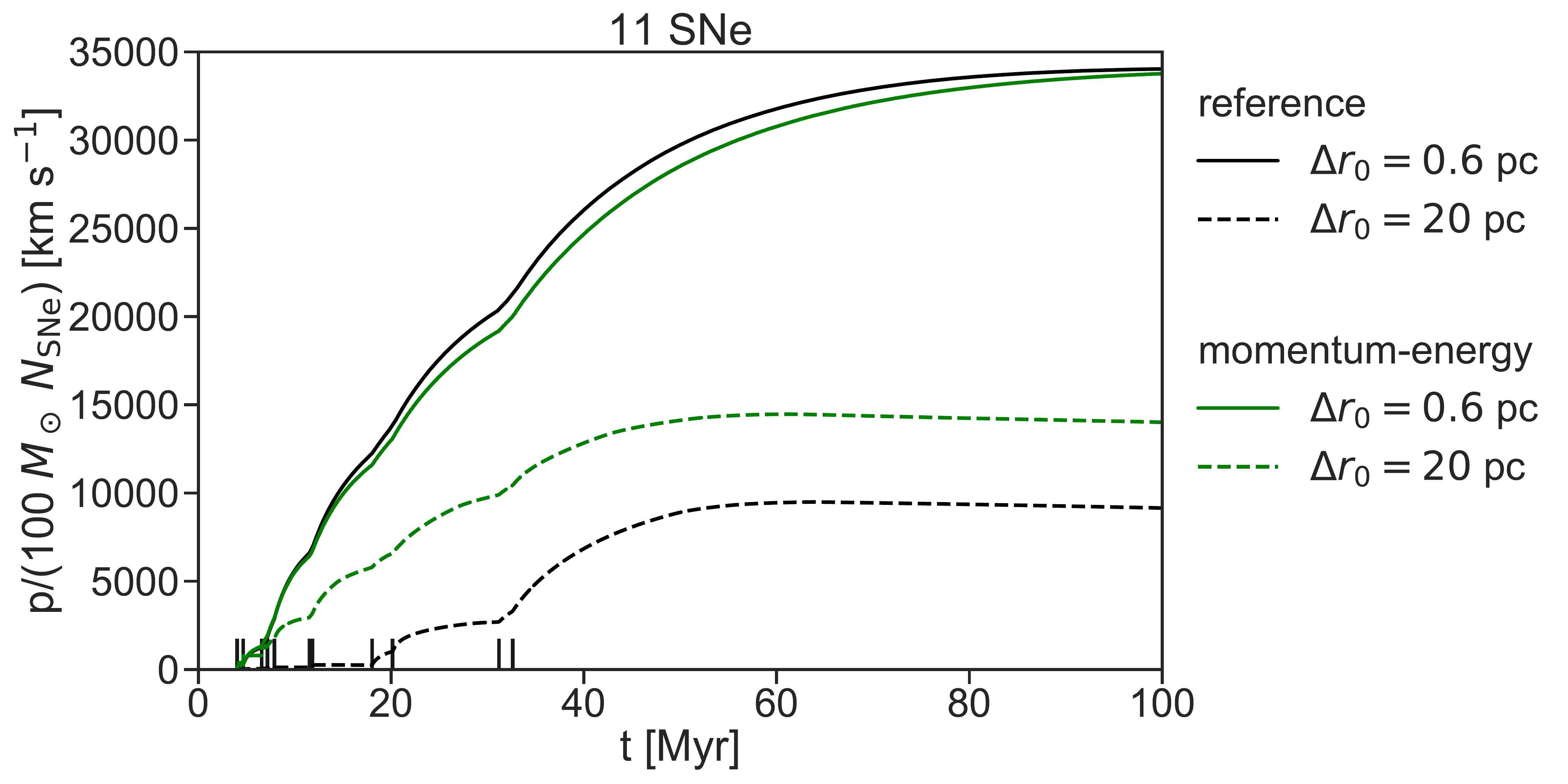}
\caption{
Same as \autoref{fig:reference:evolution}, except now overplotting the simulations using the momentum-energy feedback model. While it is difficult to see, the high resolution momentum-energy run is plotted starting at the first SN with the others; it simply matches the fiducial model so well that it is difficult to visually distinguish until $t>10$ Myr.
}
\label{fig:momentum_injection:evolution}
\end{figure}

We tested the momentum-energy injection prescription with high and low resolution simulations of 1 and 11 SN clusters. The momentum evolution is compared to our reference simulations in \autoref{fig:momentum_injection:evolution}.

The clearest discrepancy between the results of this scheme and our high resolution fiducial results is for the 1 SN, low-resolution simulation, which has a terminal momentum which is too high (by $\sim 30\%$). However, this is relatively easy to diagnose; \citet{2018MNRAS.480..800H} prescribe $\sim 5 \times 10^5$ $M_\odot$ $N_\mathrm{SNe}$ km s$^{-1}$ of momentum per SN (at $\rho = 1.33 m_\mathrm{H}$ cm$^{-3}$), so it is not surprising at that low resolution this assumption yields more than the $3 \times 10^5$ $M_\odot$ $N_\mathrm{SNe}$ km s$^{-1}$ recovered at high resolution. \citeauthor{2018MNRAS.480..800H} adopt this value from the earlier 1D simulations of \citet{1988ApJ...334..252C}, and to be consistent with the earlier \texttt{FIRE-1} simulations that used this value.
Our 1D simulations improve on those of \citet{1988ApJ...334..252C} in many ways -- for example by use of modern cooling tables and by adoption of a pseudo-Lagrangian high-order hydrodynamics method -- and thus our somewhat lower value is likely more reliable. However, for our purposes here the offset between our high resolution results and the results of the momentum-energy feedback method at low resolution are not particularly significant, since they result from a particular numerical parameter choice, which can easily be changed.

At 11 SNe, we find that the adoption of the momentum-energy injection method helps mitigate the effects of low resolution, yielding a terminal momentum that is a factor of $\approx 1.5$ higher than a na\"ive low resolution direct injection method, though still a factor of $\approx 2.5$ too small compared to the high resolution results. Although the momentum-energy injection results do not match as well as the best delayed cooling results using a value of $E_{\rm blast}$ tuned to match the 11 SN case, the momentum-energy prescription does significantly better than the delayed cooling model using the lower $E_{\rm blast}$ recommend by \citet{2006MNRAS.373.1074S}.
Furthermore, since these results already have some margin above the minimally acceptable value, it is possible that the isolated blast terminal momentum prescription (\autoref{eq:p_t_i}) could be weakened so that simulations of both 1 SN and 11 SNe clusters match our high resolution reference results. In our testing, even simply decreasing the maximum injected momentum by a factor of 2 led to little decrease in the 11 SN, low resolution simulation's final momentum; the early momentum-driven blasts appear most important for opening the bubble when energy-dominated blasts would not suffice, but then most of the momentum is actually gained after the transition to energy-dominated blasts. However, in this work since we do not test a large diversity of cluster sizes, nor do we have a specific target final momentum for our 11 SN cluster, we simply note that it might be possible to weaken the isolated SN terminal momentum prescription, but cannot offer a specific prescription that we can show to be better.

At high resolution, the momentum-energy injection model behaves very similarly to the direct injection, reference model. The first blast already starts as an energy-dominated injection event, and after that point a low density bubble is formed at the centre of the simulation, further driving down the mass and density within the blast kernel, ensuring that all subsequent blasts also remain energy dominated. With that in mind, the largest difference relative to the direct injection simulations is that the energy-dominated blasts in the momentum-energy model are assumed to be fully kinetic, whereas the direct injection, reference simulations assume fully thermal blasts, explaining the slight discrepancy in results.

\subsection{Simultaneous energy injection results}
\label{section:results:simultaneous}

\begin{figure}
\includegraphics[width=\columnwidth]{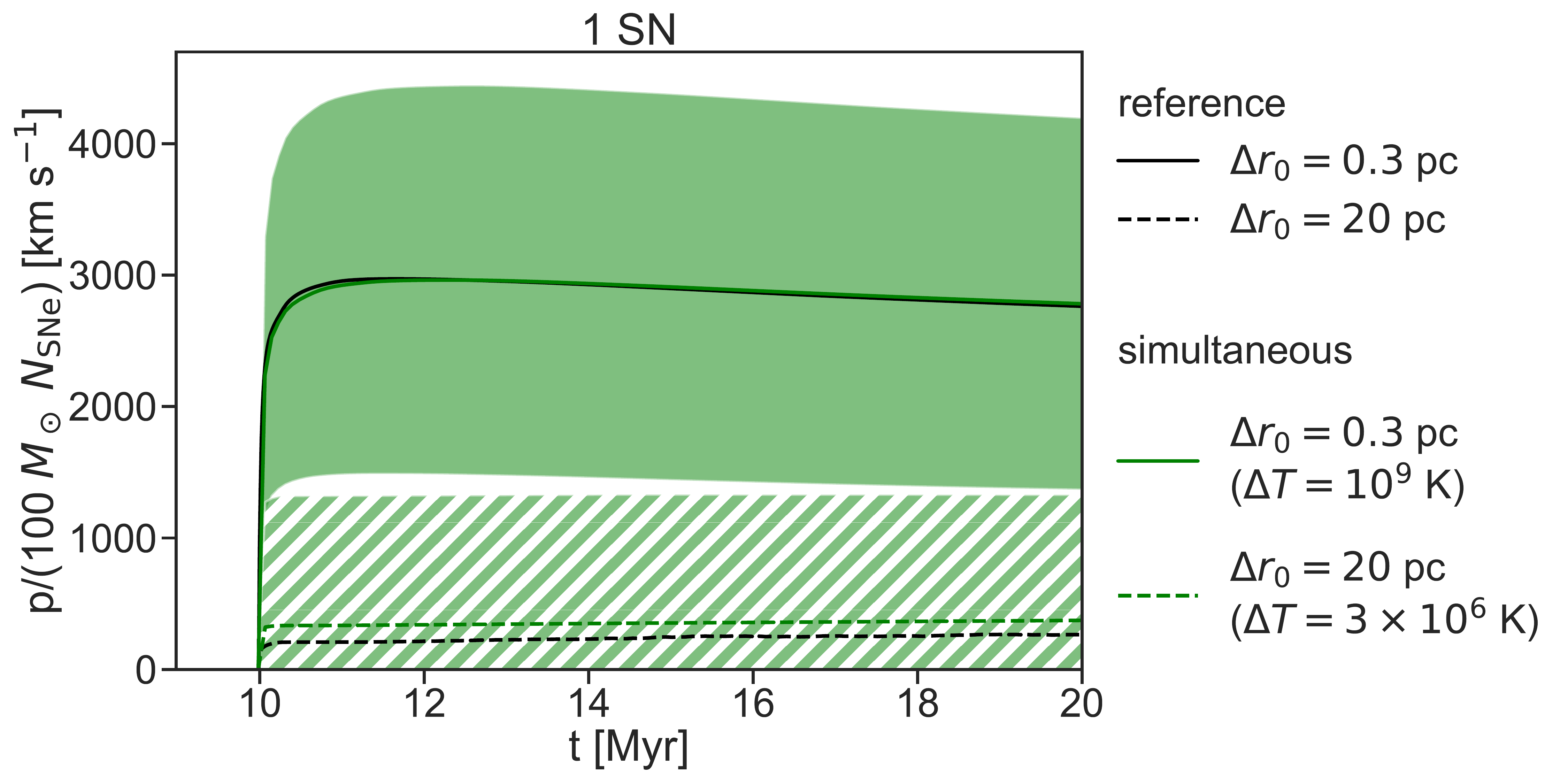}
\includegraphics[width=\columnwidth]{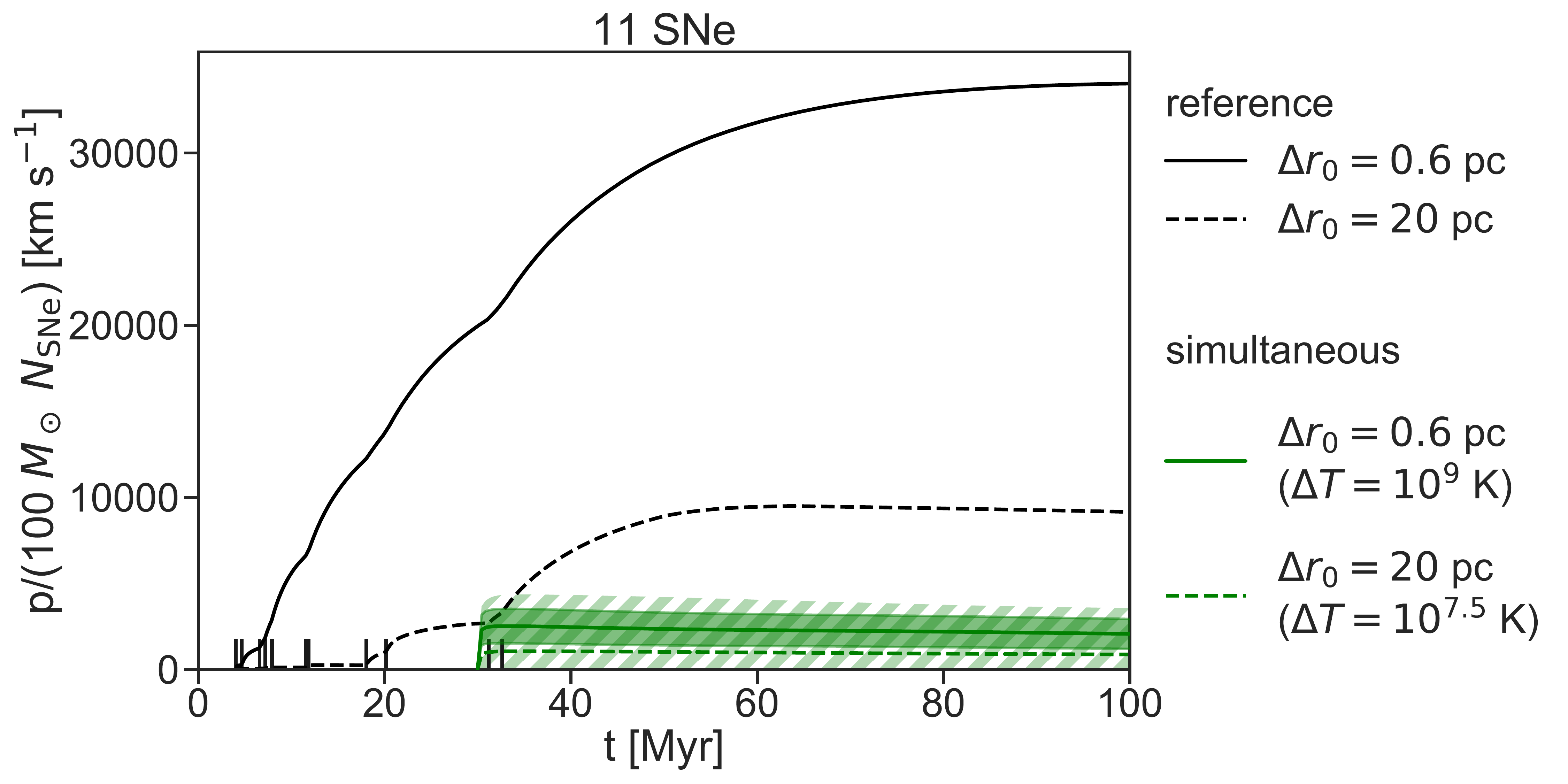}
\caption{
Same as \autoref{fig:reference:evolution}, except now overplotting the simultaneous energy injection simulations. For each simulation, we ran all eight possible realisations of the stochastic injection process. In this figure for each timestep we show the probability-weighted mean with a central green line and shade $\pm 1$ [probability-weighted] standard deviation around the mean---the high resolution run uses a solid shading, while the low resolution run uses a hatched shading.
}
\label{fig:simultaneous:evolution}
\end{figure}

\begin{figure}
\includegraphics[width=\columnwidth]{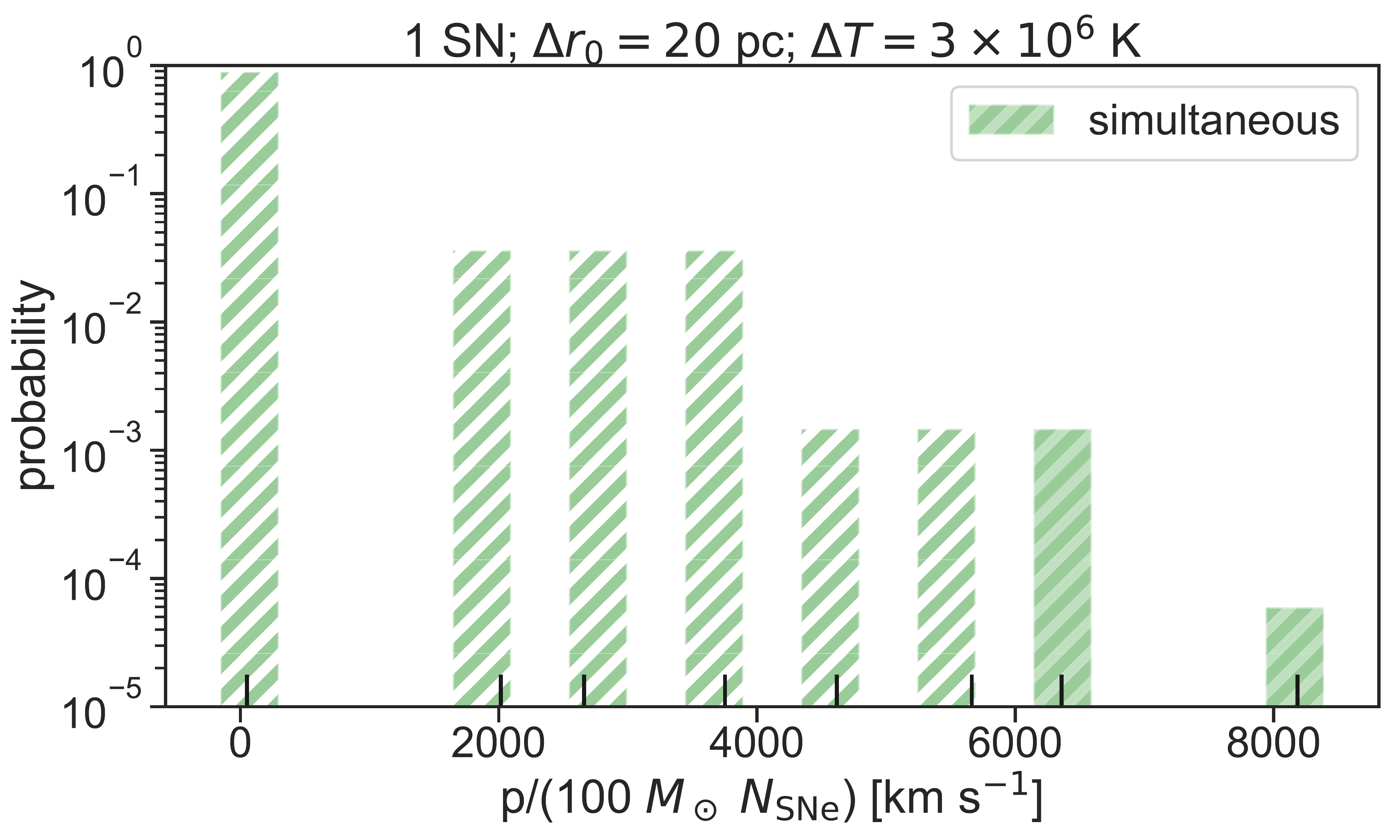}
\includegraphics[width=\columnwidth]{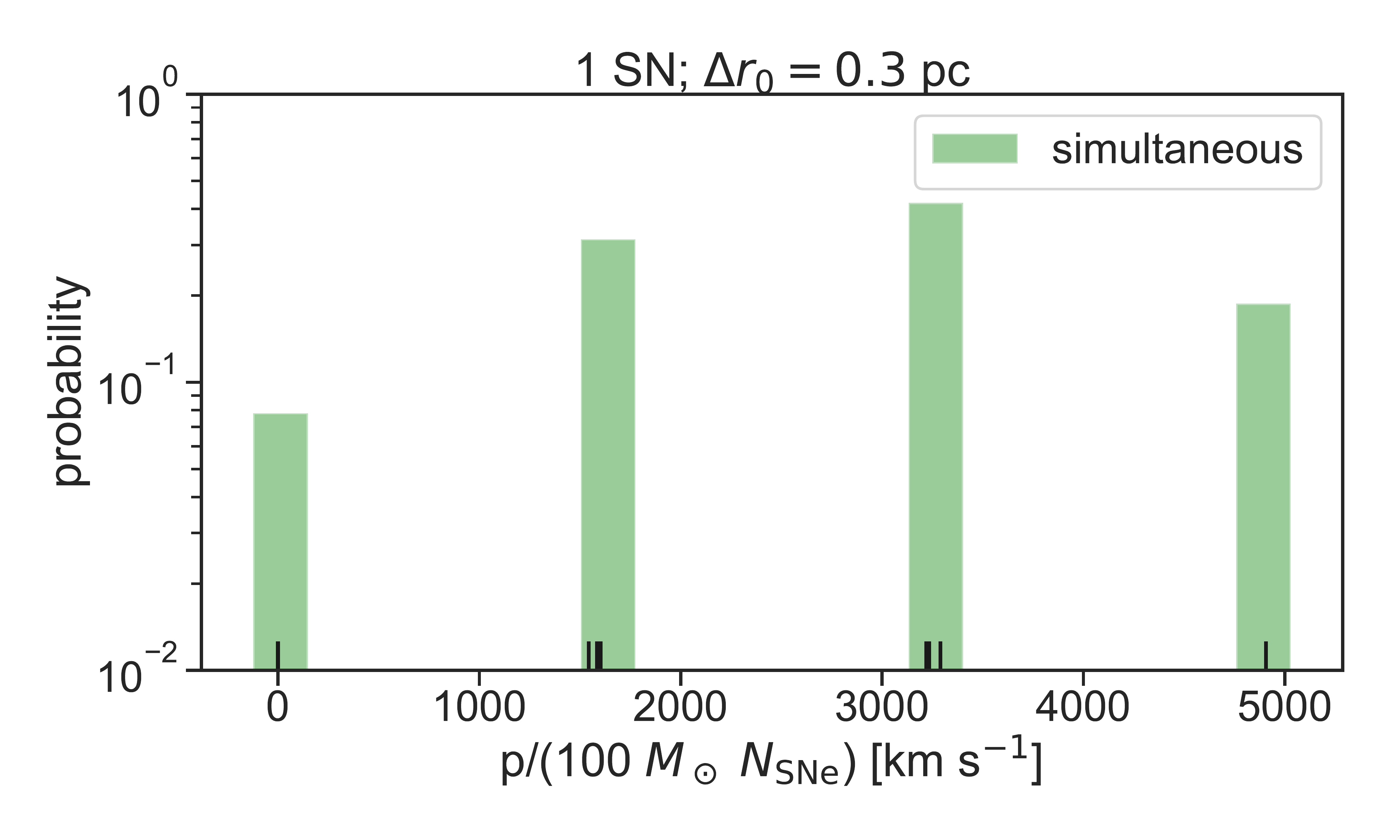}
\caption{
The final ($t= 20$ Myr) momentuma for all realisations of the 1 SN simultaneous energy injection model, with the low resolution realisations on the left and the high resolution realisations on the right. The exact momentum of each realisation is marked by a solid black tick mark; for the histogram, each realisation contributes its Bernoulli probability of occurrence to the binned probability mass function.
}
\label{fig:simultaneous:final:1}
\end{figure}

\begin{figure*}
\includegraphics[width=.49\textwidth]{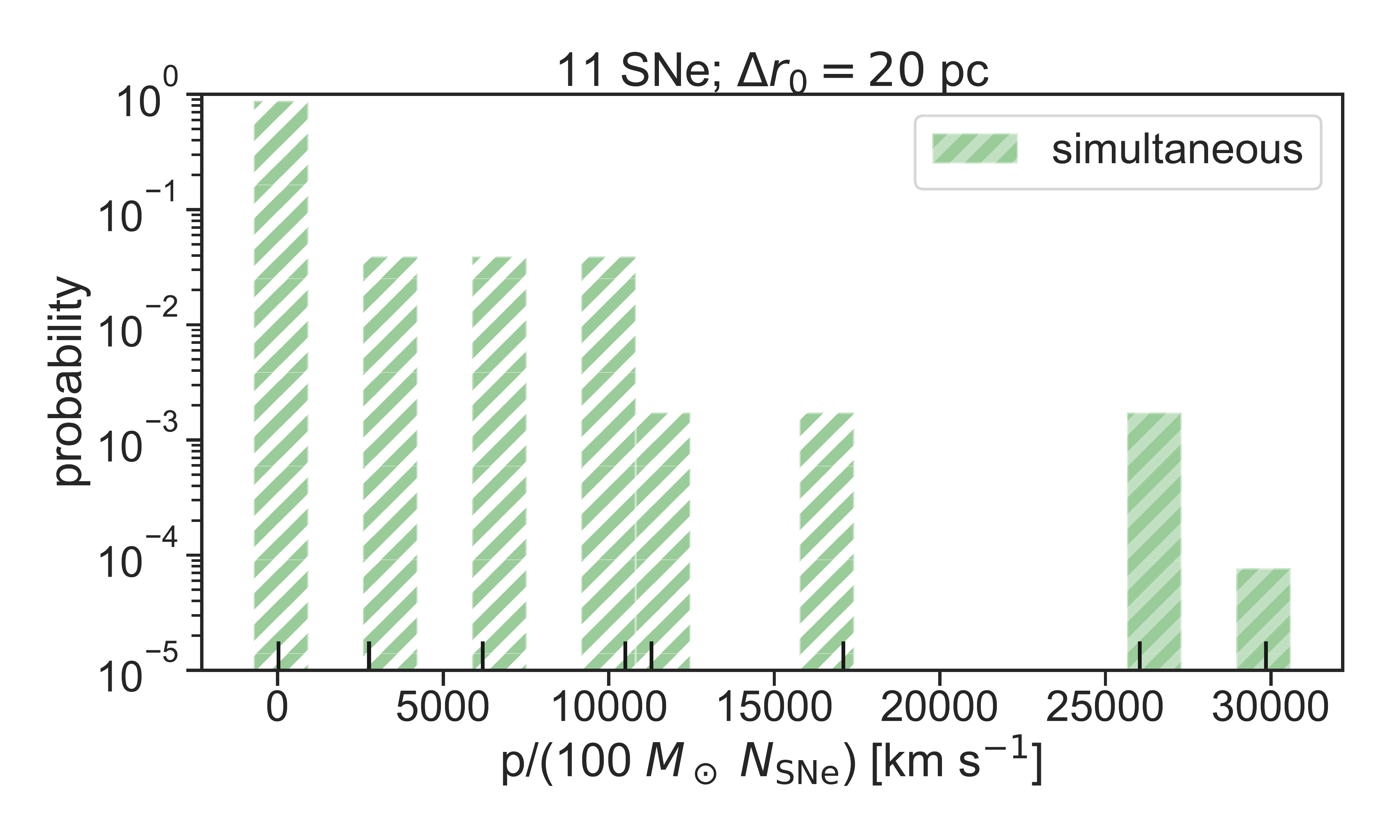}
\includegraphics[width=.49\textwidth]{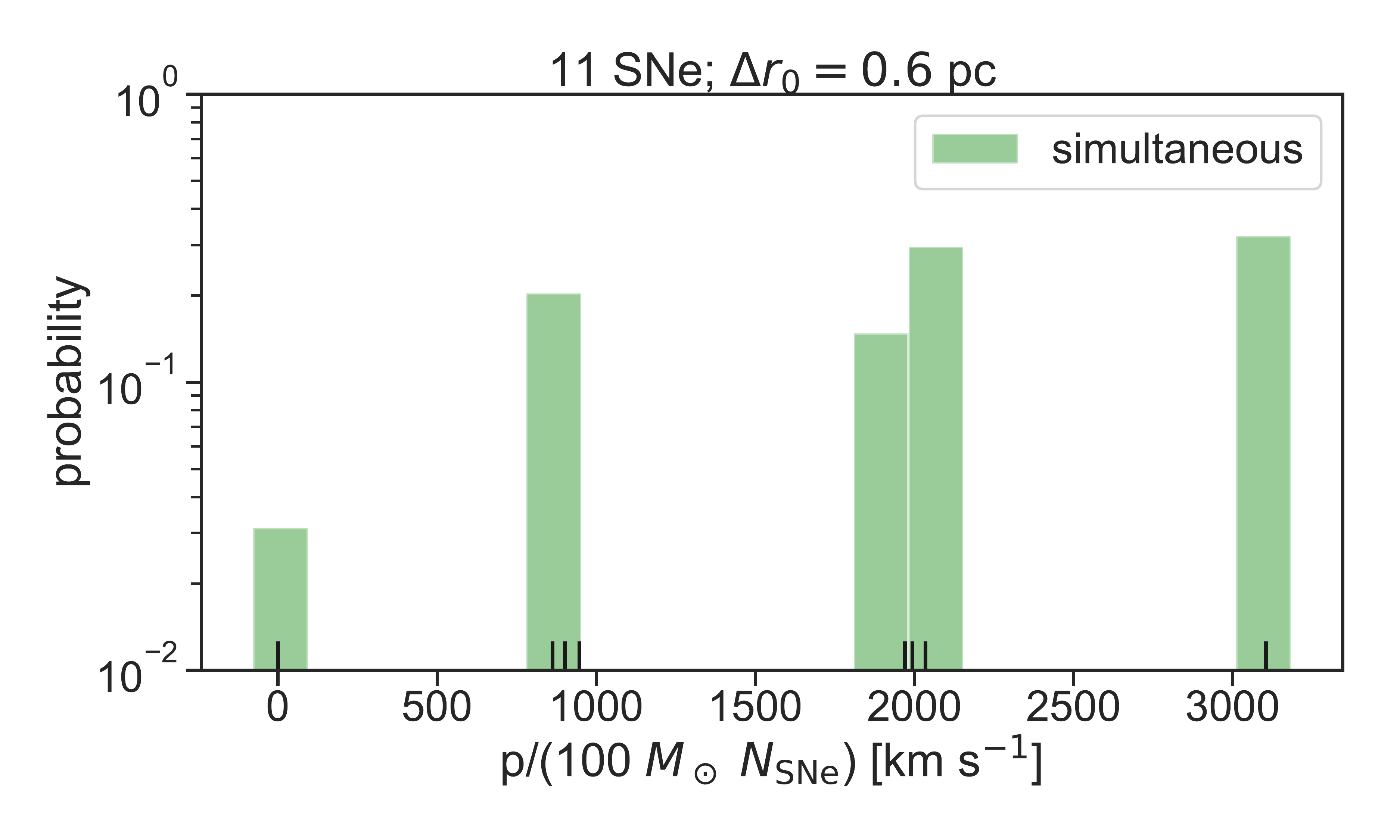}
\caption{
Same as \autoref{fig:simultaneous:final:1}, except now for our 11 SN simulations using the simultaneous injection model extracted at $t=100$ Myr.
}
\label{fig:simultaneous:final:11}
\end{figure*}

Finally, we compare the simultaneous energy injection approach to our reference simulations, with the momentum evolution shown in \autoref{fig:simultaneous:evolution}. Since there are so many realisations (8 high resolution and 8 low resolution), we simplify these figures by only showing the probability-weighted mean, along with mean $\pm$ 1 standard deviation at each point in time.  The full distribution of final momenta can be seen in Figures~\ref{fig:simultaneous:final:1} and \ref{fig:simultaneous:final:11} for the 1 SN simulations and 11 SN simulations respectively. No substantial difference was observed in the shape of the time evolution of the momentum of each realisation besides the overall normalisation.

First, for 1 SN at high resolution, we see in \autoref{tab:results} and \autoref{fig:simultaneous:evolution} that the mean momentum corresponds well to the results from our reference simulations. There is some scatter (see \autoref{fig:simultaneous:final:1}), as some realisations inject $>10^{51}$ of energy and others inject $<10^{51}$, but the probability-weighted mean result is close to the standard expected value. This is unsurprising, because for 1 SN most theoretical studies predict a roughly linear scaling between momentum and energy at high resolution (e.g., \citealt{1988ApJ...334..252C}, \citealt{2011piim.book.....D}), and the expected energy is unbiased by construction ($\langle \Delta E \rangle = E_\mathrm{blast}$). At low resolution for 1 SN, we see the simultaneous energy injection model does not do significantly better than the direct injection model, suggesting that it is still susceptible to overcooling even though we have ensured that at the time of energy injection $t_{\mathrm{cooling}} \approx 10 \times t_\text{sound crossing}$. Ultimately this suggests that it is not enough to only prevent overcooling at the initial time of injection (by ensuring a fixed $\Delta T$); overcooling in the subsequent shock front is also an important consideration.

Moving to 11 SNe we find somewhat worse results. At high resolution, we find the average momentum per SN for 11 SNe is roughly consistent with our results for 1 SN; this is bad because even our low resolution reference simulation showed a substantial momentum efficiency boost. Even when we look at the high resolution realisation which adds the most energy ($\approx 1.6 \times 10^{52}$ erg, i.e. $\approx 50\%$ above the mean value), it still results in less momentum than the 11 SN low resolution reference simulation.
This is an indication that the dynamics of a single, larger blast are different from the dynamics of a series of smaller blasts. If we na\"ively extrapolate the momentum-energy scaling relation which is roughly linear for isolated SNe (e.g., \citealt{1988ApJ...334..252C}, \citealt{2011piim.book.....D}), we would expect the mean momentum per SN to be roughly constant with respect to the number of SNe. That is qualitatively what we observe for these high resolution simultaneous energy injection simulations, whereas our reference simulations (e.g., bottom panel of \autoref{fig:reference:evolution}) clearly do not resemble a superposition of isolated blasts and instead seem to progressively build in intensity as the ISM is preprocessed by the preceding blasts.

At low resolution, while we do see a momentum efficiency boost for the 11 SN simultaneous injection realisations relative to their 1 SN simultaneous injection counterparts, it still has not clearly improved beyond the momentum efficiency of high resolution single SN simulations. This general situation is especially concerning since the cluster mass ($\approx 10^3 M_\odot$) is actually well paired to the kernel mass at low resolution ($70 \times 10^3 M_\odot$), as this implementation intended.

So in conclusion, the simultaneous energy injection model does not appear to be very effective at producing the correct final momentum. For the 1 SN cluster its performance is comparable to the direct injection model, but for the 11 SN cluster it does worse than the direct injection model.

%%%%%%%%%%%%%%%%%%%%%%%%%%%%%%%%%%%%%%%%%%%%%%

\section{Discussion}
\label{section:discussion}

\subsection{Comparison of injection methods}

In this section we focus on our primary question: which subgrid models produce a momentum efficiency for 11 SNe at low resolution that is at least a factor of 2 greater than the fiducial momentum efficiency of an isolated SN (i.e., $p_\mathrm{end} / (100 M_\odot N_\mathrm{SNe}$ km s$^{-1}) > 5500$)?   This is a necessary but not sufficient test; we predict the true momentum efficiency boost for this cluster is roughly a factor of 2, but these 1D simulations might overpredict the momentum that application of the same subgrid model in 3D would yield due to artificially suppressing asymmetries from hydrodynamic instabilities that drive mixing. Therefore, a subgrid model that cannot produce a factor of 2 boost in momentum efficiency in 1D would also necessarily underproduce momentum in 3D. However, we cannot assume the converse: if a model produces greater than a factor of 2 enhancement in momentum efficiency in 1D, it still might produce too little momentum in 3D.

Looking at \autoref{tab:results}, we can easily answer this question for the three subgrid models.
The delayed cooling model clears this threshold for $E_\mathrm{blast} = 10^{51}$ but not $E_\mathrm{blast} = 10^{50}$, the momentum-energy model clears this threshold, and the simultaneous injection model does not clear this threshold on average (although an individual realisation has a $\sim 10\%$ probability that it will clear this threshold; see \autoref{fig:simultaneous:final:11}).

As noted in Sections~\ref{section:intro} and \ref{section:results:delayedcooling}, this is not a perfect test. First, we do not precisely know the true momentum efficiency for this cluster; we can only extrapolate existing 3D simulations. 
Second, here we have only tested two cluster sizes ($N_\mathrm{SNe} = 1$ and 11), whereas we have shown that a particular choice of model parameters might perform well at one cluster scale and poorly at another (in particular, see \autoref{section:results:delayedcooling}). Therefore we cannot be sure how these models might perform across a realistic distribution of cluster sizes without a more methodical study, although by choosing an approximately maximal enhancement cluster ($N_\mathrm{SNe} = 11$), we hope to bound the problem.
Third, by using 1D simulations we are artificially suppressing asymmetries produced by hydrodynamic instabilities that increase mixing and can decrease the final momentum. For instance, our reference model at low resolution clears our factor of 2 enhancement threshold, even though it almost certainly would result in too little momentum in 3D. This illustrates that our threshold could be too lenient, but to raise it higher would require a prediction about how additional mixing would affect each subgrid model. We cannot measure that directly from these simulations for each subgrid model, but we can make qualitative predictions.

\subsection{Predicted effects of mixing}
\label{section:discussion:mixing}

Although the low resolution (20 pc), 11 SN simulation using our reference model shows a factor of $\sim 3$ increase in momentum efficiency relative to the fiducial isolated SN momentum efficiency, some 3D  simulations of the same cluster \citep{2019MNRAS.483.3647G} showed an apparent \emph{decrease} in momentum efficiency relative to the fiducial isolated value, even at much higher resolutions (e.g., 2 pc) due to the presence of increased mixing in 3D. This mixing is entirely numerical: the same simulation run at higher resolution shows a momentum increase, with the momentum continuing to increase even at the finest resolution available. Thus, while we do not have corresponding 3D simulations of this cluster for the subgrid models tested here, it is nonetheless important to consider what impact mixing might have in 3D at the resolutions likely to be typical of the simulations in which these recipes are deployed. 

\paragraph*{Delayed Cooling} 
For the 11 SN, delayed cooling simulations with $E_\mathrm{blast} = 10^{51}$ erg, no cooling can occur until the shock moves beyond $R_\mathrm{E} \approx 100$ pc or $t_\mathrm{E} \approx 7$ Myr pass. In our low resolution simulation, the radius restriction passes first, after about 1 Myr has elapsed, at which point we can observe noticeable radiative cooling. Unfortunately the momentum efficiency at this time only reached about $p(t) / (10^2 N_\mathrm{SNe} M_\odot \text{ km s}^{-1}) \sim 1500$; it has not yet achieved the isolated SN momentum efficiency let alone the desired factor of 2 enhancement. Most of the energy is still contained within radiative cooling-disabled cells, so our 1D simulation is still able to gain significant amounts of momentum beyond this time, but the same might not hold in 3D, especially if hydrodynamic instabilities are able to mix mass and energy across the interface of resolution elements with enabled and disabled radiative cooling.  Therefore it is possible that although this model performs well in 1D, it could result in too little momentum in 3D.

\paragraph*{Momentum-energy feedback}
The momentum-energy model has a useful safeguard: at low resolution, even if all the thermal energy is radiated immediately after each SN, the deposited momentum is prescribed to approximately match the isolated SN terminal momentum. This means that it is unlikely that the momentum-energy model would result in a lower momentum efficiency than the fiducial isolated SN efficiency, even in the presence of strong mixing in 3D. Furthermore, this model already prescribes slightly too much momentum (see \autoref{section:results:momentum_energy}), and if the deposited momentum is enough to open a low density bubble, then the momentum deposited with subsequent SNe can increase as the local ISM density decreases (\autoref{eq:p_t_i}). Using our low resolution single SN simulation with this model as a lower limit on the efficiency ($p/(10^2 N_\mathrm{SNe} M_\odot \text{ km s}^{-1}) \approx 4400 $), we feel confident that this model would exhibit some enhancement relative to the fiducial momentum efficiency, but cannot be sure that it would achieve the desired factor of 2 enhancement. While it might be possible to tune this model by increasing the deposited momentum per SN for larger clusters, it is unclear how to best do this for a distribution of cluster sizes and across a range of resolutions for which a bubble might or might not be opened.

\paragraph*{Simultaneous energy injection} The simultaneous energy injection model provides no direct means to prevent enhanced cooling due to mixing, but it does provide an indirect benefit: there is no time between SNe for the shock to weaken, which is when 3D instabilities like those seen by \citet{2019MNRAS.483.3647G} are especially likely to develop and strengthen.  Still, it is unlikely that this model will produce \emph{more} momentum in 3D, and given that in 1D it already produces too little momentum $\sim 90\%$ of the time, we do not expect this model would be able to capture the effect of clustering in 3D.

%%%%%%%%%%%%%%%%%%%%%%%%%%%%%%%%%%%%%%%%%%%%%%

\section{Conclusions}
\label{section:conclusions}

We set out to answer a primary question: which of the subgrid models commonly-used to model SN feedback in galactic and zoom-in cosmological-scale simulations can reproduce the factor of $>2$ increase in the terminal radial momentum delivered per SN when SNe occur in a cluster of 11 SNe rather than as single, isolated event. This question is crucial because both observations and simulations suggest that superbubbles driven by multiple SNe play an important role in regulating the formation of galactic winds and ejecting mass and metals from galaxies. Our study therefore illuminates which subgrid models can at least potentially capture this phenomenon.

The three methods we tested meet this goal with varying degrees of success. A
delayed cooling model (similar to that used in the \texttt{GASOLINE-2} code) can mimic the increase in terminal momentum of superbubbles, but only when we increase the deposited energy per SN from $E_\mathrm{blast} = 10^{50}$ erg (as recommended) to $10^{51}$ erg; the price of this choice is that it overestimates the terminal efficiency produced by a single SN. A momentum-energy model (similar to that used in the \texttt{FIRE-2} simulations) achieves this as well, without requiring any changes. A simultaneous energy injection model (similar to that used in the \texttt{EAGLE} simulations) fails with $\sim 90\%$ probability in any given stochastic realisation, as well as in the mean of the stochastic results.

This test provides a useful window into how these subgrid models behave in realistic situations, but it should be remembered that this test has limitations. First, it was done at resolutions higher than those used in very large scale cosmological simulations. This was partially because we could only fairly test resolutions at which a 11 SN cluster could plausibly be resolved, but that in itself is an indicator that star formation feedback models must take on a slightly different function when a simulation cannot even resolve the mass of a typical stellar cluster. At that point, the average output per stellar mass formed (averaged over the distribution of cluster masses) becomes more important. Fortunately our model focuses on what we think is one of the key outputs---momentum. However, that brings us to the second limitation of our test: we know simply injecting the asymptotic momentum predicted by most current models is not enough by itself. In particular, even though \citet{2018MNRAS.478..302S} found a momentum-based approach to yield the best galactic results out of a suite of simulations testing 6 feedback models, the momentum-based still failed on some key tests at low resolution (such as reproducing a realistic mass loading factor for the large scale winds). So while a momentum-energy approach might be necessary, it is not sufficient to produce accurate results at low resolutions.

Furthermore, while our results show the value of performing these tests on common SNR/superbubble regimes, they also hint at the limitations of exploring only two possible SNR regimes, as we have here.  For example, under the delayed cooling model $E_\mathrm{blast} = 10^{50}$ erg is strongly favoured over $E_\mathrm{blast} = 10^{51}$ erg for 1 SN, but for 11 SNe the opposite is true. Unfortunately, having tested only two cluster sizes, we cannot directly prescribe a solution. It is unclear if a simple switch between these two energy values, depending on whether SN are overlapping, would be sufficient to capture the main effects of clustering, or if we need a complex formula for $E_\mathrm{blast}(N_\mathrm{SNe}, \rho, Z, ...)$. As another example, with the momentum-energy model, we saw that for 1 SN at low resolution, we can directly prescribe the terminal momentum, but at 11 SNe we ended with a different, higher momentum efficiency. If this model needs to be tuned stronger or weaker to properly account for clustering, it is unclear how to do so precisely; for a single SN, we could just turn a knob, but for a superbubble more complex behaviour dynamically emerges.

Finally, in order to understand subgrid models better we would at least need to run similar tests in 3D, which is how these models are most commonly applied and where these simulations would experience stronger mixing leading to stronger cooling and lower momenta. In \autoref{section:discussion:mixing} we made qualitative predictions on how each model might be affected by mixing in 3D, but it is hard to know a quantitative value without running the experiment directly.  However, even if we ran these models in 3D (and in an ISM representative of those present in low resolution galactic simulations) our test would only be as powerful as our knowledge of the true momentum efficiency.

With regard to this last point, we remind the reader that the true momentum efficiency of clustered SN feedback is currently unknown, inherently limiting any test like this.  While we are making progress towards understanding the effects of clustering on SN feedback, discrepancies as large as a factor of 5 still exist within current literature. These discrepancies are primarily because we do not know the true level of physical mixing present in a superbubble expanding into a realistic ISM. Solving this problem likely requires converged simulations that include thermal conduction and a 3D, multi-phase, turbulent, magnetised ISM. While various simulations of clustered SNe have touched on each of these in turn, none have done so simultaneously, let alone across a range of superbubble regimes. Until that point, our ability to test, diagnose problems within and improve subgrid models of SN feedback will remain fundamentally limited.

%%%%%%%%%%%%%%%%%%%%%%%%%%%%%%%%%%%%%%%%%%%%%%%%%%%%%%%%%%%%%%%%%%%%%%%%%%%%%
\section*{Acknowledgements}

We thank the anonymous reviewer for their useful comments and suggestions.
This work was supported by the NSF through grants AST-1405962 (ESG and MRK), AST-1229745 (PM) and DGE 1339067 (ESG), by the Australian Research Council through grant FT180100375 (MRK) and by NASA through a contract to the WFIRST-EXPO Science Investigation Team (15-WFIRST15-0004), administered by the GSFC (PM). This work made use of resources and services from the National Computational Infrastructure (NCI), which is supported by the Australian Government. MRK thanks the Simons Foundation, which contributed to this work through its support for the Simons Symposium ``Galactic Superwinds: Beyond Phenomenology''.

%%%%%%%%%%%%%%%%%%%%%%%%%%%%%%%%%%%%%%%%%%%%%%%%%%%%%%%%%%%%%%%%%%%%%%%%%%%%%

\bibliographystyle{mnras}
\bibliography{feedback_models}

% \bibliography{feedback_models.bbl}

%%%%%%%%%%%%%%%%%%%%%%%%%%%%%%%%%%%%%%%%%%%%%%

\appendix

\section{Convergence in the simultaneous energy injection method}
\label{section:convergence}

As seen in Figures~\ref{fig:simultaneous:final:1} and \ref{fig:simultaneous:final:11}, even at high resolution the simultaneous energy injection method does not necessarily converge towards a deterministic result. In this section we will discuss the source of this behaviour, which touches on the limiting properties of the model with respect to several key parameters: $m_\mathrm{kernel}$, $N_\mathrm{ngb}$, $\Delta \epsilon$.  And while there are many ways we could look at the convergence of this method, we will focus on the distribution of the injected energy.

First, let's look at the mean of the distribution. By design, if $m_\mathrm{kernel} \Delta \epsilon \geq  N_\mathrm{SNe}  E_\mathrm{blast}$, then the method will be unbiased. (If that condition is not true, then the method becomes deterministic and will always inject too little energy, so we will assume an appropriate value of $\Delta \epsilon$ has been chosen for the remainder of this section.)

The more interesting quantity is the variance of the injected energy. For this method, the fractional variance is:
\begin{align}
    \mathrm{var}\left( \frac{ \Delta E }{N_\mathrm{SNe} E_\mathrm{blast}}\right) &=  \left( \frac{ m_\mathrm{kernel} \Delta \epsilon }{N_\mathrm{SNe} E_\mathrm{blast} }  - 1 \right) \sum_i \left(\frac{m_i^2}{ m_\mathrm{kernel}^2} \right) 
    \\
    \mathrm{var}\left( \frac{ \Delta E }{N_\mathrm{SNe} E_\mathrm{blast}}\right) &\approx  \left( \frac{ m_\mathrm{kernel} \Delta \epsilon }{N_\mathrm{SNe} E_\mathrm{blast}}  - 1 \right) N_\mathrm{ngb}^{-1}
\end{align}
where the second form holds for resolution elements of similar mass.

This shows that so long as $m_\mathrm{kernel} \Delta \epsilon > N_\mathrm{SNe}  E_\mathrm{blast}$, the method remains stochastic. The only way to ensure true determinism while remaining unbiased is to chose a $\Delta \epsilon$ such that $m_\mathrm{kernel} \Delta \epsilon = N_\mathrm{SNe}E_\mathrm{blast}$. While this might work for some methods (like the primary situation for which this model was envisioned: a fixed kernel mass and with an identical $N_\mathrm{SNe}$ per cluster) it does not work for all methods (such as those with varying $m_\mathrm{kernel}$ which is true for grid-based codes on which this model was also designed to work, stochastic number of SNe per clusters, or those with $m_\mathrm{kernel}$ so large that $\Delta \epsilon$ does not correspond to heating the gas beyond the peak of the cooling curve).

But there are still \emph{some} cases in which this method converges towards a deterministic result. In particular, for fixed $m_\mathrm{kernel}$, as $N_\mathrm{ngb}$ increases, the variance will approach 0. But in practice, it is typically $m_\mathrm{kernel}$ that is improved, while $N_\mathrm{ngb}$ is held fixed. We can see that that will also lead to a decrease in variance, but will eventually hit the limit $m_\mathrm{kernel} \Delta \epsilon = N_\mathrm{SNe} E_\mathrm{blast}$ if $\Delta \epsilon$ is not raised (which increases the variance). 

When designing this method to apply across a large range of scales, the best we can do is \emph{prescribe} how the variance should depend on resolution and total blast energy. For instance, if we set a minimum variance at high resolution (i.e. fixed the ratio of $m_\mathrm{kernel} \Delta \epsilon / N_\mathrm{SNe} E_\mathrm{blast}$) then the injected energy would converge to a distribution and always remain non-trivially stochastic (e.g. the right panel of \autoref{fig:simultaneous:final:1} shows how energy is distributed as a scaled binomial for our implementation). If instead we adopted a different function for $\Delta \epsilon(m_\mathrm{kernel}, N_\mathrm{SNe})$, we could force the variance to shrink to 0 as resolution increases, resulting in a delta function for the distribution of injected energy. \citet{2012MNRAS.426..140D} do not specify a recommended form for $\Delta \epsilon(m_\mathrm{kernel}, N_\mathrm{SNe})$ at high resolution, so the the exact behaviour will depend on the particular implementation.

\section{3 SN Cluster}
\label{section:3SNe}

\begin{figure}
\includegraphics[width=\columnwidth]{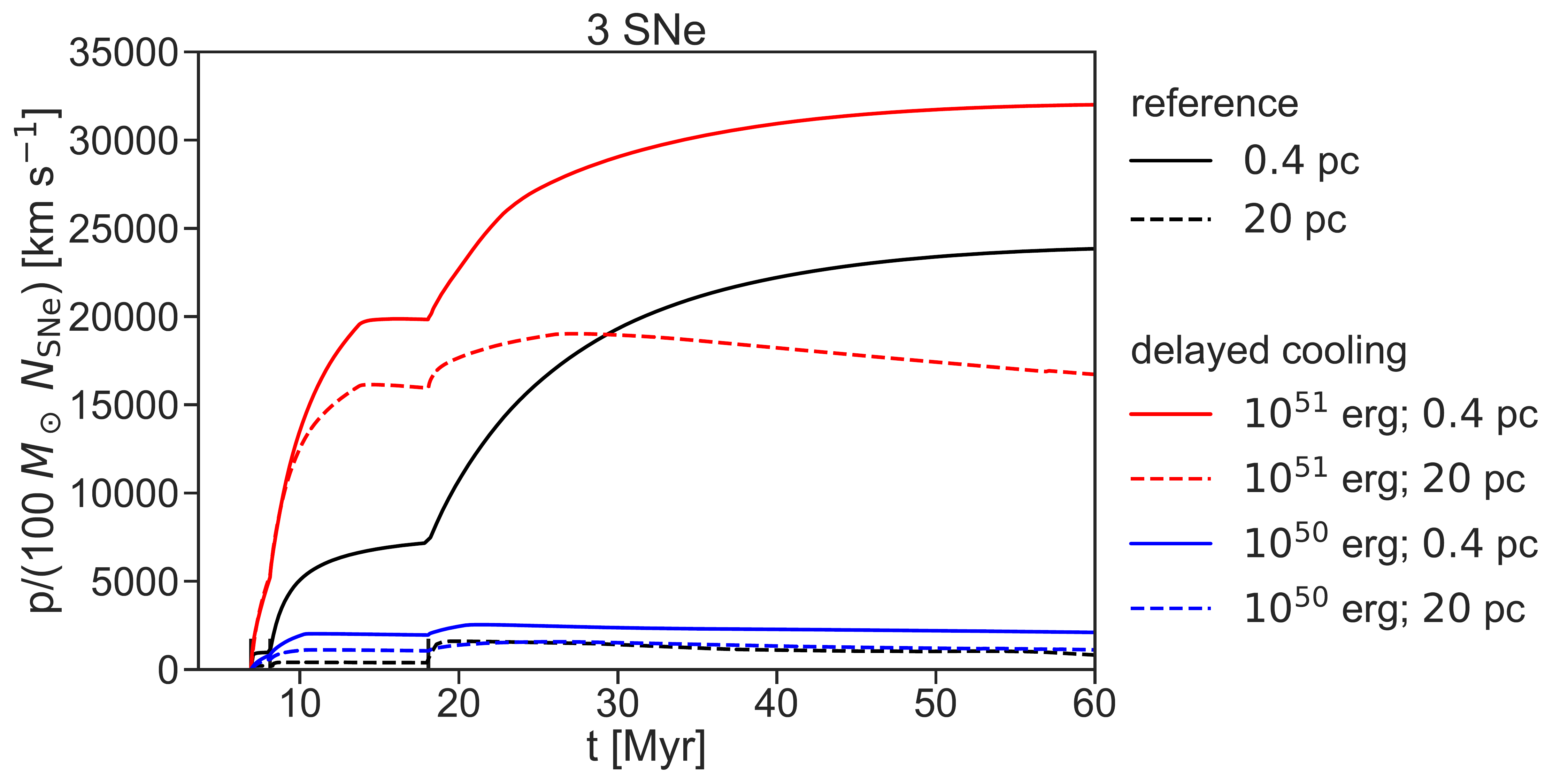}
\caption{
Same as \autoref{fig:delayedcooling:evolution}, except now for a 3 SN cluster.
}
\label{fig:delayedcooling:evolution:3SNe}
\end{figure}

\begin{figure}
\includegraphics[width=\columnwidth]{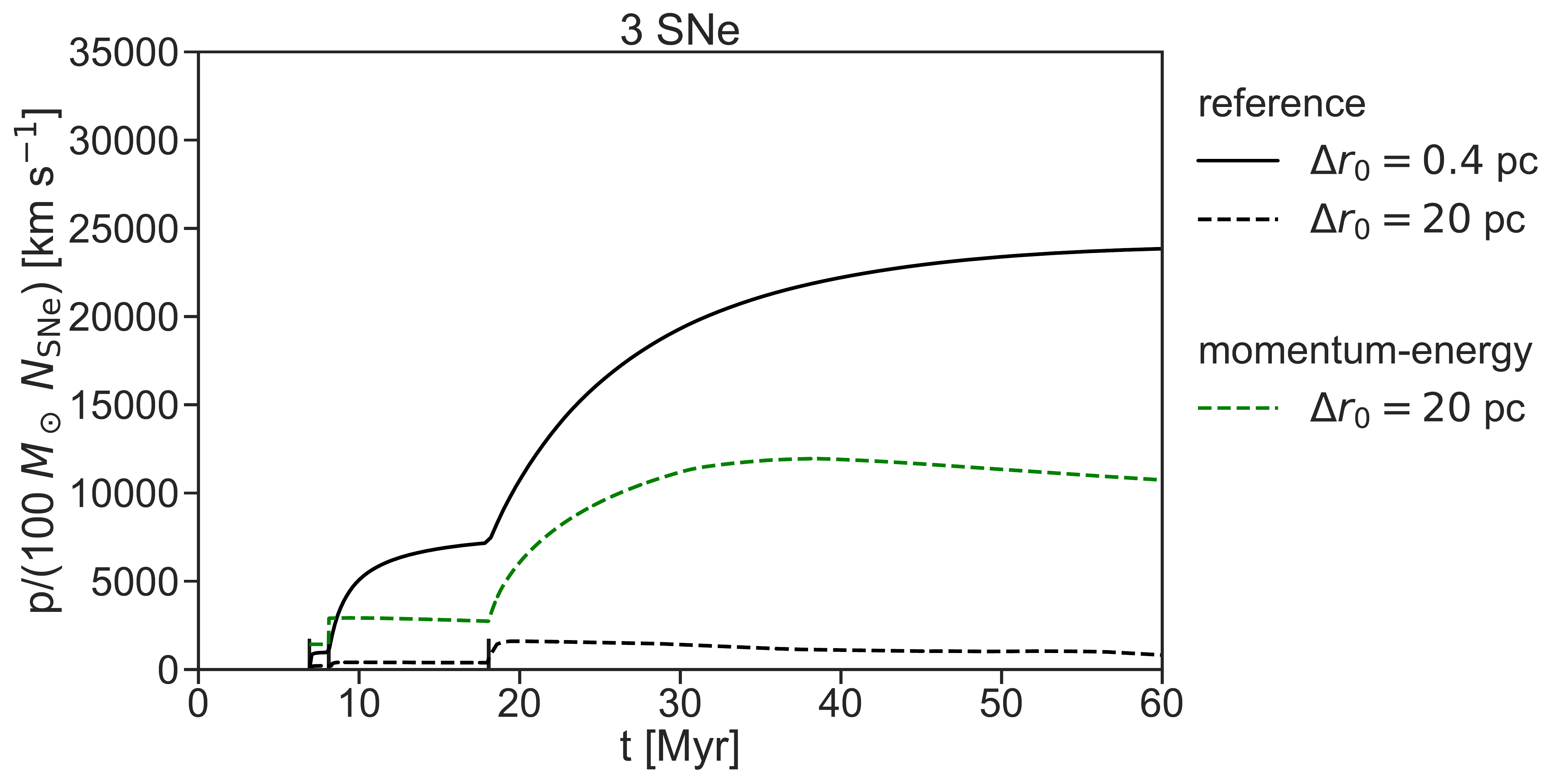}
\caption{
Same as \autoref{fig:momentum_injection:evolution}, except now for a 3 SN cluster and without a high resolution run using the momentum-energy feedback model.
}
\label{fig:momentum_injection:evolution:3SNe}
\end{figure}

\begin{figure}
\includegraphics[width=\columnwidth]{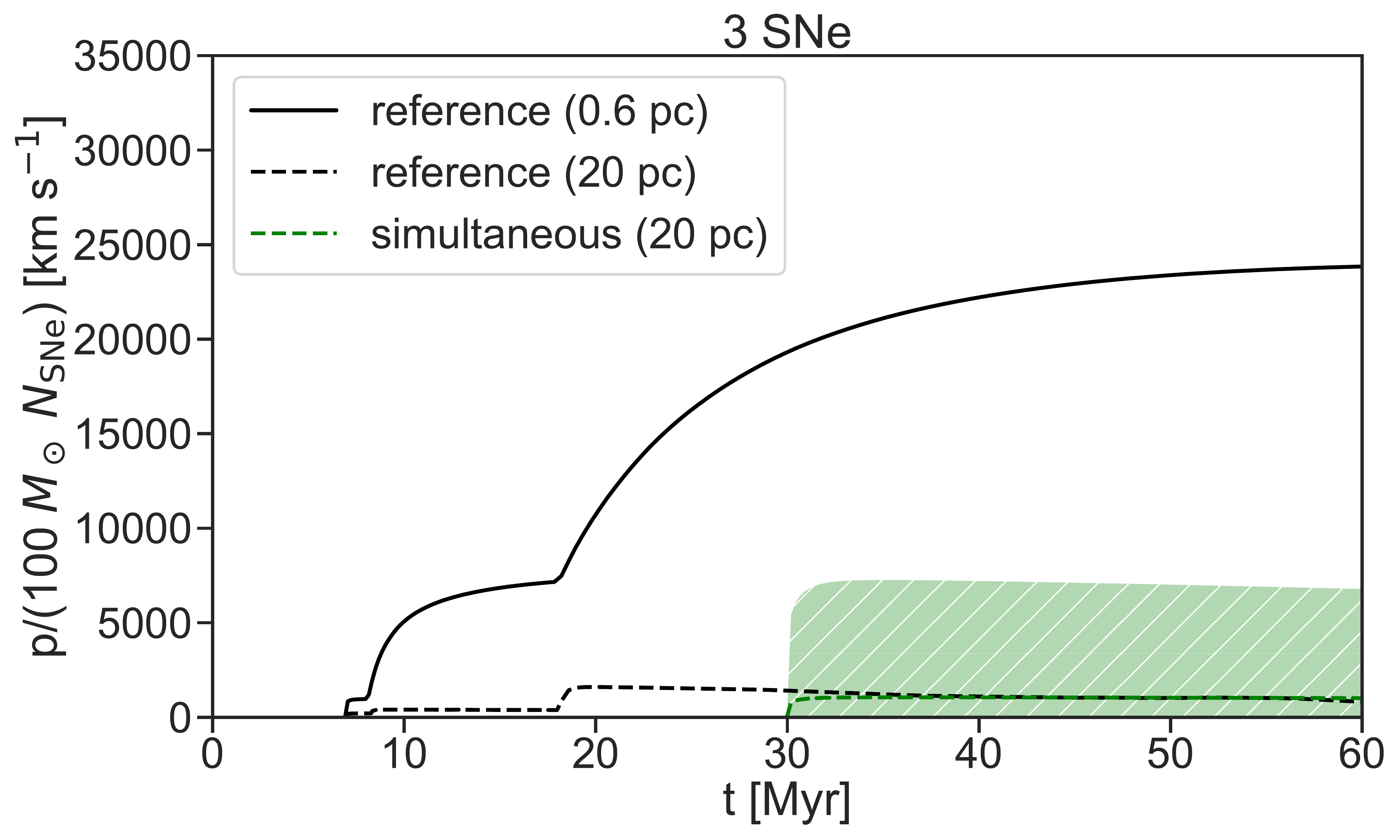}
\caption{
Same as \autoref{fig:simultaneous:evolution}, except now for a 3 SN cluster and without high resolution realisations using the simultaneous injection feedback model.
}
\label{fig:simultaneous:evolution:3SNe}
\end{figure}

To get a better understanding of how the behaviours of these feedback models are affected by the number of SNe produced by a cluster, we ran a few extra simulations of a 3 SN cluster (corresponding to an initial stellar mass of 300 $M_\odot$). The evolution of the radial momentum of these simulations is shown in Figures~\ref{fig:delayedcooling:evolution:3SNe}-\ref{fig:simultaneous:evolution:3SNe}. Since we are most interested in the performance of the feedback model simulations run at low resolution, not every feedback model was run at high resolution.

Without a 3D resolution study of this 3 SN cluster, we are unable to determine which of these models performs ``well-enough'' in the way that we could for the body of this paper. Instead, we analyse these results qualitatively.  Starting with the delayed cooling model (see \autoref{fig:delayedcooling:evolution:3SNe}), we can see that with $E_\mathrm{blast}=10^{51}$ it does not largely overpredict the momentum as it did with a 1 SN cluster but not an 11 SN cluster (see \autoref{section:results:delayedcooling}).  This means that it is possible the shutoff cooling model performs well always using $E_\mathrm{blast} = 10^{51}$ erg as long as $N_\mathrm{SNe} > 1$. However we cannot say if choosing a constant $E_\mathrm{blast}$ is preferred over a more general, unknown function $E_\mathrm{blast}(N_\mathrm{SNe})$ since we do not know the ``true'' final momentum to target. For the momentum-energy model (see \autoref{fig:momentum_injection:evolution:3SNe}), we also see behaviour that more closely matches the 11 SN cluster than the 1 SN cluster; it predicts slightly less momentum at low resolution than the reference model predicts at high resolution. Finally, on average the simulataeneous realisations (with $\Delta T = 3 \times 10^7$ K) show a \emph{decrease} in the momentum efficiency relative to the standard isolated SN value (see \autoref{fig:simultaneous:evolution:3SNe}). Since we have no 3D, 3 SN simulation, we cannot rule out the possibility that a 3 SN cluster \emph{should} show a decreased momentum efficiency, but circumstantially it is a sign that the simultaneous model likely produces too little momentum.

%%%%%%%%%%%%%%%%%%%%%%%%%%%%%%%%%%%%%%%%%%%%%%

% Don't change these lines
\bsp	% typesetting comment
\label{lastpage}
\end{document}